\documentclass[a4paper,12pt]{article}
\usepackage{jheppub}
\usepackage[T1]{fontenc}
\usepackage{graphicx,subfigure}
\usepackage{braket}
\usepackage{bm}
\usepackage{comment}

\usepackage{acro}
\DeclareAcronym{gr}{
    short=GR ,
    long=general relativity
}
\DeclareAcronym{uv}{
    short=UV ,
    long=ultraviolet
}
\DeclareAcronym{ir}{
    short=IR ,
    long=infrared
}
\DeclareAcronym{rg}{
    short=RG ,
    long=renormalization group
}
\DeclareAcronym{qg}{
    short=QG ,
    long=quantum gravity
}
\DeclareAcronym{kk}{
    short=KK ,
    long=Kaluza-Klein
}
\DeclareAcronym{qft}{
    short=QFT ,
    long=quantum field theory 
}
\DeclareAcronym{nec}{
    short=NEC ,
    long=null energy condition
}
\DeclareAcronym{anec}{
    short=ANEC ,
    long=averaged null energy condition
}
\DeclareAcronym{flrw}{
    short=FLRW ,
    long=Friedmann-Lematre-Robertson-Walker
}
\allowdisplaybreaks[1]

\title{\bf Pauli-Villars regularization of Kaluza-Klein Casimir energy with Lorentz symmetry 
}

\author{\large Hiroki Matsui ${}^a$, Yutaka Sakamura ${}^{b,c}$}

\emailAdd{hiroki.matsui@yukawa.kyoto-u.ac.jp}
\emailAdd{sakamura@post.kek.jp}
\affiliation{${}^a$Center for Gravitational Physics and Quantum Information,
Yukawa Institute for Theoretical Physics, Kyoto University,
Kitashirakawa Oiwakecho, Sakyo-ku,
Kyoto 606-8502, JAPAN\medskip\\
${}^{b}$KEK Theory Center, Institute of Particle and Nuclear Studies, KEK,
1-1 Oho, Tsukuba, Ibaraki 305-0801, Japan\medskip\\
${}^{c}$Graduate University for Advanced Studies (Sokendai),
1-1 Oho, Tsukuba, Ibaraki 305-0801, Japan\medskip\\}

\abstract{
The Pauli-Villars regularization is appropriate to discuss the UV sensitivity of low-energy observables 
because it mimics how the contributions of new particles at high energies cancel 
large quantum corrections from the light particles in the effective field theory. 
We discuss the UV sensitivity of the Casimir energy density and pressure in an extra-dimensional model 
in this regularization scheme, 
and clarify the condition on the regulator fields to preserve the Lorentz symmetry of the vacuum state. 
Some of the conditions are automatically satisfied in spontaneously-broken supersymmetric models, 
but supersymmetry is not enough to ensure the Lorentz symmetry. 
We show that the necessary regulators can be introduced as bulk fields. 
We also evaluate the Casimir energy density with such regulators, 
and its deviation from the result obtained in the analytic regularization. 
}
\begin{document} 
\begin{flushright}
{\small KEK-TH-2593 
\\YITP-23-174
}\\
\end{flushright}
\maketitle
\flushbottom

\section{Introduction}

The Casimir effect is a macroscopic quantum effect that has been observed in various experiments 
and the observed values are in good agreement with theoretical predictions~\cite{Casimir:1948dh,Lamoreaux:1996wh,Mohideen:1998iz,Roy:1999dx,Bimonte:2021sib}. 
The Casimir energy is defined as the energy difference between 
the vacuum energy in a compact space, such as a space enclosed by conducting plates, 
and that in a non-compact space. 
The vacuum energy in \ac{qft} is generally divergent and must be regularized, such as the cutoff regularization, in which the cutoff scale~$\Lambda_{\rm cut}$ is set 
for the momenta of virtual particles in the loops. 
It is well-known that the Casimir energy remains finite even in the limit of $\Lambda_{\rm cut}\to \infty$.  
The scale~$\Lambda_{\rm cut}$ is regarded as a scale at which the theory under consideration 
breaks down and is replaced with a more fundamental theory. 
The Casimir effect also plays an important role in extra-dimensional models. 
The quantum correction for the extra-dimensional models is \ac{kk} Casimir energy, which depends on the compactification scale~$m_{\rm KK}$ and determines the physical properties of the extra-dimensional models~\cite{Garriga:2000jb,Toms:2000bh,Goldberger:2000dv,Brevik:2000vt}.
Since the extra-dimensional models are non-renormalizable, they should be regarded as effective theories of more fundamental ones, such as a string theory or \ac{qg}. 
Hence $\Lambda_{\rm cut}$ can not be infinite, and 
may be close to $m_{\rm KK}$. 
In the latter case, the unknown \ac{uv} physics can affect the \ac{kk} 
Casimir energy. 
This indicates that the Casimir energy in the extra-dimensional models have 
regularization dependence~\cite{Beneventano:1995fh,Moretti:1998rf,Hagen:2000bu,
Visser:2016ddm,Matsui:2018tan,Asai:2021csl}, 
in contrast to the case of renormalizable theories, in which we can safely take the limit~$\Lambda_{\rm cut}\to\infty$. 
In particular, one of the authors suggests the 
Casimir energy receives a large correction from the \ac{uv} physics 
when $\Lambda_{\rm cut}$ is not far from $m_{\rm KK}$ in the cutoff regularization scheme~\cite{Asai:2021csl}. 

In $3+1$-dimensional \ac{qft}, there is a significant discussion regarding the Lorentz symmetry violation 
in the regularization of vacuum energy. 
Indeed, when utilizing the cutoff regularization, the \ac{uv} divergences break the Lorentz symmetry~\cite{Akhmedov:2002ts,Koksma:2011cq,Martin:2012bt}. 
If this Lorentz symmetry violation is considered as an actual physical phenomena, 
they could lead to significant cosmological issues~\cite{Danielsson:2018qpa}.
When we consider the \ac{flrw} universe with the metric 
$ds^{2}=dt^{2}-a^{2}\left(t\right)\delta_{ij}dx^{i}dx^{j}$, 
the semiclassical Friedmann equations for a flat universe with a vacuum state are given by
\begin{align}
&H^2=\frac{1}{3}\left(\Lambda_{\rm cc}+\Braket{0|\hat{\rho}|0}\right),\label{eq:Friedmann1}\\
&2\dot{H}+3H^2=\Lambda_{\rm cc}-\Braket{0|\hat{p}|0}, 
\end{align}
where 
$\Lambda_{\rm cc}$ is the cosmological constant, the hat denotes an operator,
$\hat{\rho}$ is the energy density, $\hat{p}$ is the pressure, 
the dot denotes the time derivative, and $H\equiv  \dot{a}/a$ is the Hubble parameter.
A combination of these equations leads to 
\begin{align}
 \dot{H} &=-\Braket{0|\hat{\rho}+\hat{p}|0}. 
\end{align}
In the cutoff regularization, we have 
\begin{align}
\begin{split}
\Braket{0|\hat{\rho}+\hat{p}|0}&= (-1)^{\delta_{if}}
\int_{0}^{\Lambda_{\rm cut}} \frac{d^3k}{2(2\pi)^3}\;
\left\{\sqrt{k^2+m^2}+\frac{k^2}{3\sqrt{k^2+m^2}}\right\}\\
&=(-1)^{\delta_{if}}
\frac{\Lambda_{\rm cut} ^3 \sqrt{\Lambda_{\rm cut} ^2+m^2}}{12 \pi ^2}\,,\label{eq:Friedmann2}
\end{split}
\end{align}
where $\delta_{if}$ is the Kronecker delta with $i = b, f$ for bosons and fermions respectively, and $m$ is the mass.
This shows that the \ac{uv} divergences directly contribute to the dynamics of the universe. 
\footnote{We briefly mention the observational constraints on $\Braket{0|\hat{\rho}+\hat{p}|0}$. 
These are derived from the current measurements of 
the dark energy and the constraints on its equation of state $w_{\rm dark}$. 
The Lorentz violation by dark energy can be formalized by the following expression:
\begin{equation}
\rho_{\rm dark} +p_{\rm dark} =\left(1+w_{\rm dark}\right)\rho_{\rm dark}
\sim \left(1+w_{\rm dark}\right) (10^{-3}{\rm eV})^4\,.
\end{equation}
Although some results suggest a slight phantom-like equation of state, $w_{\rm dark}\simeq -1.03$, 
several independent observations are broadly consistent with the cosmological constant value of $w_{\rm dark}=-1.013^{+0.038}_{-0.043}
$~\cite{Escamilla:2023oce}. 
Thus, the vacuum must preserve the Lorentz symmetry with the accuracy, 
$\rho_{\rm dark} +p_{\rm dark} \lesssim \mathcal{O}(10^{-2})(10^{-3}{\rm eV})^4$. }
On the other hand, we should note that \eqref{eq:Friedmann2} for the fermionic contribution clearly violates the \ac{nec}.
The \ac{nec} is known as a necessary condition to
eliminate any pathological spacetime or unphysical geometry~\cite{Hawking:1991nk,Rubakov:2014jja} and it states $T_{\mu\nu}n^{\mu}n^{\nu}\ge  0$, for any null light-like vector $n^{\mu}$. This is summarized as $\rho+p \ge  0$ for the \ac{flrw} metric. 
In the context of the vacuum energy of the quantum fields and its regularization, there exist issues related to the breaking of Lorentz symmetry and the violation of the \ac{nec}.

In this paper, we explore the \ac{kk} Casimir energy density and pressure from compact dimension. 
We particularly study the \ac{uv} sensitivity of the \ac{kk} Casimir energy. 
As we will show in the next section, the analytic regularization inherently omits the \ac{uv} contributions, 
and the cutoff regularization violates the Lorentz symmetry in the vacuum state. 
Therefore, we adopt the Pauli-Villars regularization, which effectively demonstrates the cancellation of large quantum corrections by the contributions of high-energy virtual particles in the effective field theory. 
We further specify the necessary conditions on the regulator fields to preserve Lorentz symmetry.\footnote{
See also Ref.~\cite{Visser:2016mtr}, which discusses related issues.
} 
Although spontaneously-broken supersymmetric (SUSY) models satisfy some of these conditions, 
SUSY is not enough to ensure the Lorentz invariance of the vacuum state. 

The rest of this paper is organized as follows.
In Section~\ref{sec:KK-Casimir}, we review the analytic and cutoff regularizations of the \ac{kk} Casimir energy density 
and pressure. 
We point out that these regularizations are not adequate to evaluate the \ac{uv} sensitivity of the Casimir energy preserving the Lorentz symmetry. 
In Section~\ref{sec:Pauli-Villars}, we consider the Pauli-Villars regularization to regularize the Casimir energy density, 
and provide the necessary conditions for regulator fields to preserve Lorentz symmetry.
In Section~\ref{sec:UV-dependence}, we numerically calculate the 
Casimir energy density and pressure
in the Pauli-Villars regularization, and evaluate their dependence on the \ac{uv} regulator mass scale.
In Section~\ref{sec:conclusions}, we conclude our work.

\section{Regularizations}
\label{sec:KK-Casimir}
We take the following semiclassical treatment~\cite{Birrell:1982ix}, 
which approximately combines \ac{qft}  and \ac{gr}, and is expected to be reliable 
under conditions where \ac{qg} is not important. 
We treat spacetime classically 
and use the expected value of the quantized stress-energy tensor in Einstein's equations.
Hence, the quantum effect of matter fields on 
spacetime geometry can be approximately 
described by the semiclassical equations,\footnote{
Here we have taken the unit of the gravitational constant, i.e., $8\pi G_N=1$.
}
\begin{align}\label{eq:Einstein's}
{ G }_{ \mu \nu  }
+{ \Lambda_{\rm cc}  } { g }_{ \mu\nu }=\Braket{{ T }_{ \mu\nu } },
\end{align}
where ${ G }_{\mu\nu}$ is the Einstein tensor, 
$\Lambda_{\rm cc}$ is the cosmological constant and $\Braket{{ T }_{ \mu\nu } }$ 
is the expected value of the quantum stress-energy tensor. 
Phenomenologically, such treatment will suffice.
\footnote{
This approach has challenges. Specifically, the quantized stress-energy tensor in curved spacetime introduces higher-derivative corrections, leading to non-unitary massive ghosts and potential instability in spacetime and its perturbations, as referenced in various studies~\cite{Buchbinder:1992rb,Stelle:1976gc,Horowitz:1978fq,Horowitz:1980fj,Hartle:1981zt,RandjbarDaemi:1981wd,Jordan:1987wd,Suen:1988uf,Suen:1989bg,Anderson:2002fk,Matsui:2019tlf}. These quantum effects could contradict current observations if they significantly influence the universe~\cite{Matsui:2019tlf}. Thus, the semiclassical gravity may not hold up under higher-perturbative calculations and may require specialized analysis methods within the effective field theory~\cite{Simon:1991bm,Parker:1993dk}. In this paper, we do not consider such higher-order calculations.}

It is known that the (quantum) vacuum is Lorentz invariant to a high accuracy from the observation~\cite{Will:2005va,Will:2014kxa}. 
Therefore, the vacuum energy density~$\hat{\rho}$ must give rise to an energy-momentum tensor in the 4D Minkowski spacetime 
of the form, 
\begin{align}\label{T^vac}
 \Braket{T^{\rm vac}_{\mu\nu}} &= \Braket{0|\hat{\rho}|0}\eta_{\mu\nu}, 
\end{align}
where $\eta_{\mu\nu}={\rm diag}\,(-1,1,1,1)$ is the Minkowski metric, 
and thus the quantum correction to the vacuum energy density is renormalized by the cosmological constant~$\Lambda_{\rm cc}$. 
Note that \eqref{T^vac} indicates that 
\begin{align}\label{rhoplusp}
 \Braket{0|\hat{\rho}+\hat{p}|0} = 0, 
\end{align}
where $\hat{p}\equiv T^{\rm vac}_{11}=T^{\rm vac}_{22}=T^{\rm vac}_{33}$ is the vacuum pressure. 
Therefore, the LHS of \eqref{rhoplusp} measures the violation of the Lorentz symmetry.

\subsection{Formal expressions for energy density and pressure}

To simplify the discussion, we consider a real scalar theory 
in a flat 5-dimensional spacetime,
and one of the spatial dimensions is compactified on $S^1/Z_2$.  
\begin{equation}
 \mathcal{L}= -\frac{1}{2}\partial^\mu\Phi\partial_\mu\Phi-\frac{1}{2}M_{\rm bulk}^2\Phi^2, 
\end{equation}
where $\mu=0,1,\cdots,4$, and $M_{\rm bulk}$ is a bulk mass parameter. 
The coordinate of the compact dimension is denoted as $y\equiv x^4$. 
The fundamental region of $S^1/Z_2$ is chosen as $0\leq y\leq \pi R$, 
where $R$ is the radius of $S^1$. 
The real scalar field~$\Phi$ is assumed to be $Z_2$ odd. 
Then the \ac{kk} masses are given by
\begin{equation}
 m_n = \sqrt{M_{\rm bulk}^2+\frac{n^2}{R^2}}.\;\;\;\;\; \left(n=1,2,\cdots\right)
\end{equation}

The vacuum energy density and the vacuum pressure in the 4D effective theory are formally expressed as
\begin{align}\label{eq:vacuum-energy-density}
 \Braket{0|\hat{\rho}|0} &= \sum_{n=1}^\infty \int\frac{d^3k}{2(2\pi)^3}\;\sqrt{k^2+m_n^2}, \nonumber\\
 \Braket{0|\hat{p}|0} &= \frac{1}{3}\sum_{n=1}^\infty \int\frac{d^3k}{2(2\pi)^3}\;\frac{k^2}{\sqrt{k^2+m_n^2}}. 
\end{align}
These obviously diverge, and we need to regularize them. 
In the following, we review the analytic and the momentum-cutoff regularizations, 
and mention unsatisfactory points for our purpose. 
To make the relation between them clear, we introduce the cutoff for the KK mode number~$N_{\rm cut}$, 
the momentum cutoff~$\Lambda_{\rm cut}$ and the complexified dimension~$d$. 
Then, (\ref{eq:vacuum-energy-density}) is regularized as
\begin{align}
 \Braket{0|\hat{\rho}|0} &= \sum_{n=1}^{N_{\rm cut}} \int_0^{\Lambda_{\rm cut}}\frac{d^dk}{2(2\pi)^d\mu^{d-3}}\;\sqrt{k^2+m_n^2}, \nonumber\\
 \Braket{0|\hat{p}|0} &= \frac{1}{d}\sum_{n=1}^{N_{\rm cut}} \int_0^{\Lambda_{\rm cut}}\frac{d^dk}{2(2\pi)^d\mu^{d-3}}\;\frac{k^2}{\sqrt{k^2+m_n^2}},
\end{align}
where $\mu$ is some scale to adjust the mass dimension. 
Naively, the cutoff scales for the 3D momentum and the fifth one are expected to be common. 
Thus we assume that $m_{N_{\rm cut}}\simeq \Lambda_{\rm cut}$, or more specifically 
\begin{align}\label{def:N_cut}
 N_{\rm cut} &= {\rm floor}\left(R\sqrt{\Lambda_{\rm cut}^2-M_{\rm bulk}^2}\right).
\end{align}
Performing the $\vec{k}$-integral, we obtain
\begin{align}\label{regularized_expressions}
 \Braket{0|\hat{\rho}|0} &= \sum_{n=1}^{N_{\rm cut}}
 \frac{m_n^{d+1}}{2(4\pi)^{d/2}\mu^{d-3}\Gamma(\frac{d}{2})}B_{1-\epsilon_n}\left(\frac{d}{2},-\frac{d+1}{2}\right), \nonumber\\
 \Braket{0|\hat{p}|0} &= \sum_{n=1}^{N_{\rm cut}}
 \frac{m_n^{d+1}}{2d(4\pi)^{d/2}\mu^{d-3}\Gamma(\frac{d}{2})}B_{1-\epsilon_n}\left(\frac{d+2}{2},-\frac{d+1}{2}\right), 
\end{align}
where $\Gamma(\alpha)$ is the Euler gamma function, $B_z(\alpha,\beta)$ is the incomplete beta function, and
\begin{align}
 \epsilon_n &\equiv \frac{m_n^2}{\Lambda_{\rm cut}^2+m_n^2}. 
\end{align}

\subsection{Analytic regularization}

\subsubsection{Review of conventional derivation}
The most popular regularization scheme for the calculation of the Casimir energy is 
the combination of the dimensional regularization and the zeta-function regularization, 
which we call analytic regularization in this paper. 

Let us take the limit~$\Lambda_{\rm cut}\to\infty$, i.e., $\epsilon_n\to 0$, keeping $d-3$ nonzero, in (\ref{regularized_expressions}). 
Then, the incomplete beta function reduces the complete beta function, and becomes $n$-independent. 
\begin{align}\label{limit_incompleteBs}
 \lim_{\epsilon_n\to 0}B_{1-\epsilon_n}\left(\frac{d}{2},-\frac{d+1}{2}\right) 
 &= B\left(\frac{d}{2},-\frac{d+1}{2}\right) = -\frac{\Gamma(\frac{d}{2})\Gamma(-\frac{d+1}{2})}{2\sqrt{\pi}}, \nonumber\\
 \lim_{\epsilon_n\to 0}B_{1-\epsilon_n}\left(\frac{d+2}{2},-\frac{d+1}{2}\right)
 &= B\left(\frac{d+2}{2},-\frac{d+1}{2}\right) = \frac{d\Gamma(\frac{d}{2})\Gamma(-\frac{d+1}{2})}{2\sqrt{\pi}}.
\end{align}
Thus, (\ref{regularized_expressions}) become
\begin{align}\label{limit_regularized_expressions}
 \Braket{0|\hat{\rho}|0} &= \frac{\mu^4}{2(4\pi)^{d/2}\Gamma(\frac{d}{2})}B\left(\frac{d}{2},-\frac{d+1}{2}\right)
 \sum_{n=1}^\infty \left(\frac{m_n}{\mu}\right)^{d+1}  \nonumber\\
 &= -\frac{\mu^4\Gamma(-\frac{d+1}{2})}{2(4\pi)^{\frac{d+1}{2}}}\sum_{n=1}^\infty \left(\frac{m_n}{\mu}\right)^{d+1}, \nonumber\\
 \Braket{0|\hat{p}|0} &= \frac{\mu^4}{2d(4\pi)^{d/2}\Gamma(\frac{d}{2})}B\left(\frac{d+2}{2},-\frac{d+1}{2}\right)
 \sum_{n=1}^\infty \left(\frac{m_n}{\mu}\right)^{d+1} \nonumber\\
 &= \frac{\mu^4\Gamma(-\frac{d+1}{2})}{2(4\pi)^{\frac{d+1}{2}}}\sum_{n=1}^\infty \left(\frac{m_n}{\mu}\right)^{d+1}. 
\end{align}
The infinite sum over the \ac{kk} modes is evaluated 
by the zeta-function regularization technique~\cite{Goldberger:2000dv,Leseduarte:1996ah,Leseduarte:1996xr}. 
Using the formula~\eqref{def:S_0} with \eqref{int-K} in Appendix, 
the energy density is expressed as 
\begin{align}\label{eq:rho-z-regularization}
 \Braket{0|\hat{\rho}|0} &= -\frac{\mu^{3-d}\Gamma(-\frac{d+1}{2})}{2(4\pi)^{\frac{d+1}{2}}R^{d+1}}\sum_{n=1}^\infty 
 \left(\bar{M}_{\rm bulk}^2+n^2\right)^{\frac{d+1}{2}}
 \nonumber\\
 &= \frac{\mu^{3-d}M_{\rm bulk}^{d+1}\Gamma(-\frac{d+1}{2})}{4(4\pi)^{(d+1)/2}}
 -\frac{\mu^{3-d}\Gamma(-\frac{d+2}{2})}{8(4\pi)^{d/2}}RM_{\rm bulk}^{d+2} \nonumber \\
 &\quad -\frac{\mu^{3-d}M_{\rm bulk}^{\frac{d+2}{2}}}{(2\pi)^{d+1}R^{d/2}}
 \sum_{n=1}^\infty n^{-\frac{d+2}{2}}K_{\frac{d+2}{2}}\left( 2\pi n \bar{M}_{\rm bulk}\right), 
\end{align}
where $\bar{M}_{\rm bulk}\equiv RM_{\rm bulk}$, and $K_\alpha(z)$ is the modified Bessel function of the second kind. 
The first term diverges as $d\to 3$, but it does not depend on $R$ and is irrelevant to the stabilization of the extra dimension. 
Thus we simply neglect it. 
We require that the vacuum energy density in the decompactified limit~$R\to\infty$ vanishes~\cite{Kay:1978zr}. 
Thus the Casimir energy density, which is a function of $R$, is defined as
\begin{align}\label{def:rho_Casimir}
 \frac{\Braket{0|\hat{\rho}|0}_{\rm Casimir}(R)}{\pi R} &\equiv 
 \frac{\Braket{0|\hat{\rho}|0}(R)}{\pi R}-\lim_{R\to\infty}\frac{\Braket{0|\hat{\rho}|0}(R)}{\pi R}. 
\end{align}
Note that the subtraction should be performed for the 5D energy density 
since the second term is the quantity in the decompactified limit. 
Then, the second term in (\ref{eq:rho-z-regularization}) is cancelled, and we obtain 
\begin{align}\label{eq:rho-z-regularization2}
 \Braket{0|\hat{\rho}|0}_{\rm Casimir} &= -\frac{\mu^{3-d}M_{\rm bulk}^{\frac{d+2}{2}}}{(2\pi)^{d+1}R^{d/2}}
 \sum_{n=1}^\infty n^{-\frac{d+2}{2}}K_{\frac{d+2}{2}}\left( 2\pi n RM_{\rm bulk}\right) \nonumber\\
 &\to -\frac{M_{\rm bulk}^{\frac{5}{2}}}{16\pi^4R^{\frac{3}{2}}}
 \sum_{n=1}^\infty n^{-\frac{5}{2}}K_{\frac{5}{2}}(2\pi nRM_{\rm bulk}). 
\end{align}
We have taken the limit~$d\to 3$ at the last step. 
Similarly, the vacuum pressure is calculated as
\begin{align}\label{eq:p-z-regularization2}
 \Braket{0|\hat{p}|0}_{\rm Casimir} &\equiv \Braket{0|\hat{p}|0}-R\lim_{R\to\infty}\frac{\Braket{0|\hat{p}|0}}{R} \nonumber\\
 &= \frac{\mu^{3-d}M_{\rm bulk}^{\frac{d+2}{2}}}{(2\pi)^{d+1}R^{d/2}}
 \sum_{n=1}^\infty n^{-\frac{d+2}{2}}K_{\frac{d+2}{2}}\left( 2\pi n RM_{\rm bulk}\right) \nonumber\\
 &\to \frac{M_{\rm bulk}^{\frac{5}{2}}}{16\pi^4R^{\frac{3}{2}}}
 \sum_{n=1}^\infty n^{-\frac{5}{2}}K_{\frac{5}{2}}(2\pi nRM_{\rm bulk}). 
\end{align}

In the massless case $M_{\rm bulk}=0$, \eqref{eq:rho-z-regularization2} reduces to the well-known form, 
\begin{align}\label{massless_limit}
\Braket{0|\hat{\rho}|0}_{\rm Casimir}
 &= -\frac{\mu^{3-d}\Gamma(\frac{d+2}{2})\zeta(d+2)}
 {2^{d+2}\pi^{\frac{3}{2}d+2}R^{d+1}} \to -\frac{3\zeta(5)}{128\pi^6R^4}, 
\end{align}
where \(\zeta(s)\) is the Riemann zeta function. 

From \eqref{eq:rho-z-regularization2} and \eqref{eq:p-z-regularization2}, we can see that the sum of the \ac{kk} Casimir
energy density and pressure are exactly zero,
\begin{equation}
\Braket{0|\hat{\rho}+\hat{p}|0}_{\rm Casimir}=0\,.
\end{equation}
Thus, the Lorentz symmetry and NEC are both preserved in this regularization.\footnote{
We can already see this in the formal expressions in \eqref{limit_regularized_expressions}. 
}

\subsubsection{Cutoff sensitivity in analytic regularization}\label{UVsensitivity_analytic}
Although the formula~\eqref{eq:rho-z-regularization2} or \eqref{eq:p-z-regularization2} is useful 
because of its rapid convergent property, 
the analytic continuation processes make it difficult 
to see how the divergent terms are removed. 
It is well-known that this regularization only captures the logarithmic divergences, 
and is insensitive to the power-law divergences of $\Lambda_{\rm cut}$. 
To see the situation, 
let us review the procedure we have performed in \eqref{limit_incompleteBs} in more detail.  
As long as $\Lambda_{\rm cut}$ is kept finite, the incomplete beta functions in \eqref{regularized_expressions} are well-defined 
for any values of the dimension~$d$. 
Before taking the limit~$\Lambda_{\rm cut}\to\infty$, let us consider a case that $d<-1$. 
Then, using (\ref{expand:incomplete_beta}) in the Appendix, the incomplete beta functions are expanded as
\begin{align}\label{expansion_incompleteBs}
 B_{1-\epsilon_n}\left(\frac{d}{2},-\frac{d+1}{2}\right) &= B\left(\frac{d}{2},-\frac{d+1}{2}\right)
 +\frac{2}{d+1}\epsilon_n^{-\frac{d+1}{2}}-\frac{d-2}{d-1}\epsilon_n^{\frac{1-d}{2}} \nonumber\\
 &\quad
 +\frac{(d-2)(d-4)}{d-3}\epsilon_n^{\frac{3-d}{2}}+{\cal O}\left(\epsilon_n^{\frac{5-d}{2}}\right), \nonumber\\
 B_{1-\epsilon_n}\left(\frac{d+2}{2},-\frac{d+1}{2}\right) &= B\left(\frac{d+2}{2},-\frac{d+1}{2}\right)
 +\frac{2}{d+1}\epsilon_n^{-\frac{d+1}{2}}-\frac{d}{d-1}\epsilon_n^{\frac{1-d}{2}} \nonumber\\
 &\quad
 +\frac{d(d-2)}{d-3}\epsilon_n^{\frac{3-d}{2}}+{\cal O}\left(\epsilon_n^{\frac{5-d}{2}}\right). 
\end{align}
Since all the powers in RHS are positive for $d<-1$, we can safely take the limit~$\Lambda_{\rm cut}\to\infty$ 
(i.e., $\epsilon_n\to 0$), and drop all $\epsilon_n$-dependent terms. 
After dropping them, we can move $d$ to a value close to 3. 
This is what we have done in \eqref{limit_incompleteBs}. 
However, if we keep the $\epsilon_n$-dependent terms when we move $d$ to a value close to 3, 
the second and the third terms in RHS of \eqref{expansion_incompleteBs} have negative powers, 
and correspond to the quartic and quadratic divergences, respectively.\footnote{
Besides, the fourth terms also diverge as $d\to 3$ and contain logarithmic divergent terms. 
} 
Therefore, what we have done in \eqref{limit_incompleteBs} is just dropping the quartic and quadratic divergent terms {\it by hand}. 

A similar prescription has been performed when we apply the zeta-function regularization for the infinite sum over the KK modes. 
If we keep the cutoff~$\Lambda_{\rm cut}$ finite, 
the incomplete functions in \eqref{regularized_expressions} depend on the KK level~$n$, 
and cannot be factored out from the summation over $n$. 
Therefore, it is not easy to perform the exact calculation of \eqref{regularized_expressions}. 
Hence we investigate the following expression instead. 
\begin{align}\label{expr:damping}
 \Braket{0|\hat{\rho}|0} &= \sum_{n=1}^\infty \frac{m_n^{d+1}}{2(4\pi)^{d/2}\mu^{d-3}\Gamma(\frac{d}{2})}
 B_{1-\epsilon_n}\left(\frac{d}{2},-\frac{d+1}{2}\right)e^{-a^2n^2}, 
\end{align}
where $a\equiv 1/N_{\rm cut}$ is a tiny positive constant. 
Instead of the sharp cutoff at $n=N_{\rm cut}$, we introduce the damping factor~$e^{-a^2n^2}$, 
which suppresses the contribution of heavy KK modes with $m_n>\Lambda_{\rm cut}$.\footnote{
To simplify the discussion, we approximate $a$ as $a=(\Lambda_{\rm cut}R)^{-1}$. 
} 
Then, \eqref{expr:damping} is rewritten as
\begin{align}
 \Braket{0|\hat{\rho}|0} &= \frac{1}{2(4\pi)^{d/2}\mu^{d-3}R^{d+1}\Gamma(\frac{d}{2})}
 U\left(\frac{d}{2},-\frac{d+1}{2};\bar{M}_{\rm bulk}^2\right), 
\end{align}
where $U(\alpha,\beta;M^2)$ is defined in \eqref{def:U} in Appendix. 
According to the expression~\eqref{expr:U} with \eqref{def:Us} and \eqref{expr:U1}, 
this has the following terms. 
\begin{align}\label{divergent_terms}
 U\left(\frac{d}{2},-\frac{d+1}{2};\bar{M}_{\rm bulk}^2\right)
 &=  \frac{(\Lambda_{\rm cut}R)^{d+1}}{d(d+1)}-\frac{\bar{M}_{\rm bulk}^2(\Lambda_{\rm cut}R)^{d-1}}{2d(d-1)}
 +\frac{3\bar{M}_{\rm bulk}^4(\Lambda_{\rm cut}R)^{d-3}}{8d(d-3)} \nonumber\\
 &\quad
 -\frac{C_1(-\frac{d+1}{2};\bar{M}_{\rm bulk}^2)}{d\sqrt{\pi}}(\Lambda_{\rm cut}R)^{d+2}
 -\frac{C_2(-\frac{d+1}{2};\bar{M}_{\rm bulk}^2)}{d\sqrt{\pi}}(\Lambda_{\rm cut}R)^d \nonumber\\
 &\quad
 -\frac{C_3(-\frac{d+1}{2};\bar{M}_{\rm bulk}^2)}{d\sqrt{\pi}}(\Lambda_{\rm cut}R)^{d-2} 
 +\tilde{U}_2(d;\bar{M}_{\rm bulk}^2)+\cdots, 
\end{align}
where $C_i(\beta;M^2)$ ($i=1,2,3$) are defined in \eqref{def:Cs}, and
\begin{align}
 \tilde{U}_2(d;\bar{M}_{\rm bulk}^2) &\equiv 
 \frac{2(\Lambda_{\rm cut}R)^d}{d}\sum_{n=1}^\infty
 \sqrt{(\Lambda_{\rm cut}R)^2+\bar{M}_{\rm bulk}^2+n^2}\exp\left(-\frac{n^2}{(\Lambda_{\rm cut}R)^2}\right), 
\end{align}
and the ellipsis denotes terms that appeared in \eqref{eq:rho-z-regularization} 
and irrelevant terms that will vanish in the limit of $\Lambda_{\rm cut}\to\infty$ when $d=3$. 
In the limit of $d\to 3$ keeping $\Lambda_{\rm cut}$ finite, 
the terms shown in \eqref{divergent_terms} represent power-law divergent terms up to quintic in $\Lambda_{\rm cut}$. 
This is expected because we are considering the 5D theory. 
In the derivation of \eqref{eq:rho-z-regularization}, we have taken the limit of $\Lambda_{\rm cut}\to\infty$ 
for $d<-2$, where all terms shown in \eqref{divergent_terms} vanish. 
However, this treatment is equivalent to just dropping those terms {\it by hand}. 
Therefore, the analytic regularization is inappropriate for studying the UV sensitivity of the Casimir energy density or pressure.

\subsection{Cutoff regularization} \label{cutoff_reg}
Next, we consider the cutoff regularization. 
Take the limit~$d\to 3$, keeping $\Lambda_{\rm cut}$ finite, in \eqref{regularized_expressions}. 
Then we obtain 
\begin{align}
 \Braket{0|\hat{\rho}|0} &= \sum_{n=1}^{N_{\rm cut}}\frac{m_n^4}{8\pi^2}B_{1-\epsilon_n}\left(\frac{3}{2},-2\right) \nonumber\\
 &= \sum_{n=1}^{N_{\rm cut}}\left\{\frac{\Lambda_{\rm cut}\sqrt{\Lambda_{\rm cut}^2+m_n^2}(2\Lambda_{\rm cut}^2+m_n^2)}
 {32\pi^2}-\frac{m_n^4}{32\pi^2}\ln\frac{\Lambda_{\rm cut}+\sqrt{\Lambda_{\rm cut}^2+m_n^2}}{m_n}\right\}, 
 \nonumber\\
 \Braket{0|\hat{p}|0} &= \sum_{n=1}^{N_{\rm cut}}\frac{m_n^4}{24\pi^2}B_{1-\epsilon_n}\left(\frac{5}{2},-2\right) \nonumber\\
 &= \sum_{n=1}^{N_{\rm cut}}\left\{\frac{\Lambda_{\rm cut}\sqrt{\Lambda_{\rm cut}^2+m_n^2}(2\Lambda_{\rm cut}^2-3m_n^2)}{96\pi^2}
 +\frac{m_n^4}{32\pi^2}\ln\frac{\Lambda_{\rm cut}+\sqrt{\Lambda_{\rm cut}^2+m_n^2}}{m_n}\right\}. 
\end{align}
The sum of the vacuum energy density and pressure is
\begin{align}\label{eq:vacuum-energy-pressure}
 \Braket{0|\hat{\rho}+\hat{p}|0} = \sum_{n=1}^{N_{\rm cut}}\frac{\Lambda_{\rm cut}^3 \sqrt{\Lambda_{\rm cut}^2+m_n^2}}{12 \pi ^2},
\end{align}
where the logarithmic terms exactly cancel 
but the cut-off divergences remain. 
Namely, the Lorentz symmetry is violated in this regularization. 
If $\Phi$ is replaced with a 5D fermion, an overall minus sign appears in the above expressions. 
Hence \ac{nec} is also violated in that case.  

For a light mode with $m_n\ll \Lambda_{\rm cut}$, its contribution to the vacuum energy and the vacuum pressure 
can be expanded as 
\begin{align}\label{def:tlrhop}
 \tilde{\rho}(m_n) &\equiv \frac{m_n^4}{8\pi^2}B_{1-\epsilon_n}\left(\frac{3}{2},-2\right) \nonumber\\
 &= \frac{1}{16\pi^2}\left(\Lambda_{\rm cut}^4+\Lambda_{\rm cut}^2m_n^2+\frac{m_n^4}{8}\right)
 -\frac{m_n^4}{32\pi^2}\ln\frac{2\Lambda_{\rm cut}}{m_n}+{\cal O}\left(\frac{m_n^6}{\Lambda_{\rm cut}^2}\right), \nonumber\\
 \tilde{p}(m_n) &\equiv \frac{m_n^4}{24\pi^2}B_{1-\epsilon_n}\left(\frac{5}{2},-2\right) \nonumber\\
 &= \frac{1}{48\pi^2}\left(\Lambda_{\rm cut}^4-\Lambda_{\rm cut}^2m_n^2-\frac{7m_n^4}{8}\right)
 +\frac{m_n^4}{32\pi^2}\ln\frac{2\Lambda_{\rm cut}}{m_n}+{\cal O}\left(\frac{m_n^6}{\Lambda_{\rm cut}^2}\right). 
\end{align}
After summing over the KK modes, the leading terms of $\Braket{0|\hat{\rho}|0}$ and $\Braket{0|\hat{p}|0}$ are
\begin{align}
 \Braket{0|\hat{\rho}|0}_{\rm leading} &\sim \frac{N_{\rm cut}\Lambda_{\rm cut}^4}{16\pi^2} 
 \sim \frac{R\Lambda_{\rm cut}^5\sqrt{1-M_{\rm bulk}^2/\Lambda_{\rm cut}^2}}{16\pi^2}, \nonumber\\
 \Braket{0|\hat{p}|0}_{\rm leading} &\sim \frac{N_{\rm cut}\Lambda_{\rm cut}^4}{48\pi^2} 
 \sim \frac{R\Lambda_{\rm cut}^5\sqrt{1-M_{\rm bulk}^2/\Lambda_{\rm cut}^2}}{48\pi^2}, 
\end{align}
where $N_{\rm cut}$ is defined in (\ref{def:N_cut}). 
Since these are proportional to $R$, they are canceled in the Casimir energy density and pressure 
defined in \eqref{def:rho_Casimir} and \eqref{eq:p-z-regularization2}, respectively. 
However, the other terms remain and violate the Lorentz symmetry. 
Thus the cutoff regularization is considered to be problematic 
for the calculation of the Casimir energy~\cite{Akhmedov:2002ts,Koksma:2011cq,Martin:2012bt}. 

From the physical point of view, contributions of massive KK modes near the cutoff scale~$\Lambda_{\rm cut}$ 
should be suppressed by UV physics. 
In the previous work~\cite{Asai:2021csl}, we introduced a damping function, such as
\begin{align}
 g_{\rm damp}(n) &= \exp\left(-\frac{n^2}{2N_{\rm cut}^2}\right), 
\end{align}
or
\begin{align}\label{kink_damp}
 g_{\rm damp}(n) &= \frac{1}{2}\left[1+\tanh\left\{A\left(1-\frac{n}{N_{\rm cut}}\right)\right\}\right], 
\end{align}
where $A\gtrsim 10$ is a positive constant that controls the steepness around the cutoff scale, 
and inserted it into the expression~\eqref{regularized_expressions} as
\begin{align}
 \Braket{0|\hat{\rho}|0} &= \sum_{n=1}^\infty\frac{m_n^{d+1}}{2(4\pi)^{d/2}\Gamma(\frac{d}{2})}
 B_{1-\epsilon_n}\left(\frac{d}{2},-\frac{d+1}{2}\right)g_{\rm damp}(n). 
\end{align}
Then we obtain a finite value for the Casimir energy density, 
which agrees with the value obtained by \eqref{eq:p-z-regularization2}.\footnote{
With the damping function in (\ref{kink_damp}), the parameter~$A$ has to be chosen to a value 
in the appropriate region to obtain a consistent value with \eqref{eq:p-z-regularization2}. 
}
The cutoff regularization considered in this subsection corresponds to the limit of $A\to\infty$ in \eqref{kink_damp}. 
It is known that the regularization with such a sharp cutoff provides a divergent Casimir energy density, 
and should not be applied to the calculations for the Casimir energy density and pressure~\cite{Asai:2021csl}.

\section{Pauli-Villars regularization}
\label{sec:Pauli-Villars}
As mentioned in Sec.~\ref{cutoff_reg}, the contributions of massive KK modes near $\Lambda_{\rm cut}$ should be 
suppressed by the UV physics, such as contributions of new particles with masses of ${\cal O}(\Lambda_{\rm cut})$. 
Such contributions can be mimicked by the Pauli-Villars regulators. 
However, a single regulator that has opposite statistics and a large mass~$M_{\rm reg}$ is not enough 
to suppress contributions of the KK modes heavier than $M_{\rm reg}$.\footnote{
In the Pauli-Villars regularization, $M_{\rm reg}$ plays a role of $\Lambda_{\rm cut}$ in the cutoff regularization. 
} 
Hence, for each KK mode with mass~$m_n$, we introduce $k$ species of regulators. 
Then, its contributions to the Casimir energy density and pressure are modified as 
\begin{align}
 \rho_n &\equiv \tilde{\rho}(m_n)-\sum_{i=1}^kc_i\tilde{\rho}(M_i), \nonumber\\
 p_n &\equiv \tilde{p}(m_n)-\sum_{i=1}^kc_i\tilde{p}(M_i), 
\end{align}
where $\tilde{\rho}$ and $\tilde{p}$ are defined in (\ref{def:tlrhop}), 
$M_i$ and an integer~$c_i$ denote the mass and  the degree of freedom for the $i$-th regulator, respectively. 
We assume that all $M_i$ ($i=1,2,\cdots,k$) are of ${\cal O}(M_{\rm reg})$. 
Note that we introduce both bosonic ($c_i<0$) and fermionic ($c_i>0$) regulators.
Then, using the expanded expressions in (\ref{def:tlrhop}), we have 
\begin{align}\label{PVregulated_rhop}
 \rho_n &= \frac{1}{16\pi^2}\left\{\left(1-\sum_ic_i\right)\Lambda_{\rm cut}^4
 +\Lambda_{\rm cut}^2\left(m_n^2-\sum_ic_iM_i^2\right)+\frac{1}{8}\left(m_n^4-\sum_ic_iM_i^4\right)\right\} \nonumber\\
 &\quad
 -\frac{m_n^4}{32\pi^2}\ln\frac{2\Lambda_{\rm cut}}{m_n}
 +\sum_ic_i\frac{M_i^4}{32\pi^2}\ln\frac{2\Lambda_{\rm cut}}{M_i}
 +{\cal O}\left(\frac{M_{\rm reg}^6}{\Lambda_{\rm cut}^2}\right), \nonumber\\
 p_n &= \frac{1}{48\pi^2}\left\{\left(1-\sum_ic_i\right)\Lambda_{\rm cut}^4
 -\Lambda_{\rm cut}^2\left(m_n^2-\sum_ic_iM_i^2\right)-\frac{7}{8}\left(m_n^4-\sum_ic_iM_i^4\right)\right\} \nonumber\\
 &\quad
 +\frac{m_n^2}{32\pi^2}\ln\frac{2\Lambda_{\rm cut}}{m_n}-\sum_ic_i\frac{M_i^4}{32\pi^2}\ln\frac{2\Lambda_{\rm cut}}{M_i}
 +{\cal O}\left(\frac{M_{\rm reg}^6}{\Lambda_{\rm cut}^2}\right). 
\end{align}
If we require the integers~$c_i$ ($i=1,2,\cdots,k$) to satisfy~\cite{Koksma:2011cq,Visser:2016mtr}
\footnote{
Wolfgang Pauli found these constraints~\eqref{c:constraints}.}
\begin{align}\label{c:constraints}
 \sum_{i=1}^kc_i &= 1, \;\;\;\;\;
 \sum_{i=1}^kc_iM_i^2 = m_n^2, \;\;\;\;\;
 \sum_{i=1}^kc_iM_i^4 = m_n^4, 
\end{align}
the Lorentz-violating terms are canceled, and obtain
\begin{align}
 \rho_n &= -\frac{m_n^4}{32\pi^2}\ln\frac{M_{\rm reg}}{m_n}+\sum_{i=1}^kc_i\frac{M_i^4}{32\pi^2}\ln\frac{M_{\rm reg}}{M_i}, \nonumber\\
 p_n &= \frac{m_n^4}{32\pi^2}\ln\frac{M_{\rm reg}}{m_n}-\sum_{i=1}^kc_i\frac{M_i^4}{32\pi^2}\ln\frac{M_{\rm reg}}{M_i}, 
\end{align}
in the limit of $\Lambda_{\rm cut}\to \infty$. 
Hence we have 
\begin{align}
 \rho_n+p_n &= 0. 
\end{align}

The first condition in \eqref{c:constraints} is the requirement of the balance 
between the bosonic and fermionic degrees of freedom. 
The second one has the same form as the supertrace mass formula 
in a model that has spontaneously broken supersymmetry (SUSY)~\cite{Ferrara:1979wa}. 
Namely, the first two conditions in \eqref{c:constraints} are automatically satisfied 
in such a model. 
To preserve the Lorentz symmetry, however, the third condition is also necessary. 
It is intriguing to discuss the possibility of constructing a SUSY model in which all conditions in (\ref{c:constraints}) are satisfied~\cite{Visser:2016mtr}.

To suppress the contributions of the massive KK modes heavier than $M_{\rm reg}$, 
we should also require that 
\begin{align}\label{lim:rhop}
 \lim_{n\to\infty}\rho_n &= \lim_{n\to\infty}p_n = 0.
\end{align}
This is rewritten as
\begin{align}\label{fourth_constraint}
 \lim_{n\to\infty}\left(
 m_n^4\ln\frac{m_n^2}{M_{\rm reg}^2}-\sum_{i=1}^kc_iM_i^4\ln\frac{M_i^2}{M_{\rm reg}^2}\right)
 &= 0. 
\end{align}

In the case of
\begin{align}\label{parameter_choice}
 k &= 3, \;\;\;\;\;
 c_1 = 1, \;\;\;\;\;
 c_2 = -c_3, 
\end{align}
we can solve \eqref{c:constraints}, and obtain 
\begin{align}\label{parameter_choice:k3}
 &M_2^2 = \frac{c_2-1}{2c_2}M_1^2+\frac{c_2+1}{2c_2}m_n^2, \nonumber\\
 &M_3^2 = \frac{c_2+1}{2c_2}M_1^2+\frac{c_2-1}{2c_2}m_n^2. 
\end{align}
For $c_2=3$, we have 4 bosonic and 4 fermionic degrees of freedom in total 
and can be embedded into a chiral multiplet in a (spontaneously broken) SUSY model. 
In the following, we consider the case of \eqref{parameter_choice} with $c_2=3$ as a specific example. 

We assume that $M_i^2$ ($i=1,2,3$) are functions of $m_n^2$ and $M_{\rm reg}^2$. 
In solving \eqref{fourth_constraint}, we are interested in the KK modes with $m_n\gg M_{\rm reg}$. 
Thus, we expand $M_1^2$ as
\begin{align}
 M_1^2 &= \alpha m_n^2\left(1+\beta_1\delta+\beta_2\delta^2+\cdots\right), 
\end{align}
where $\delta\equiv M_{\rm reg}^2/m_n^2$. 
Using this expression and \eqref{parameter_choice:k3}, we can expand the LHS in \eqref{fourth_constraint} as 
\begin{align}
 &\quad
 m_n^4\ln\frac{m_n^2}{M_{\rm reg}^2}-\sum_{i=1}^3c_iM_i^4\ln\frac{M_i^2}{M_{\rm reg}^2} \nonumber\\
 &= \left({\cal C}_1m_n^4+{\cal C}_2m_n^2M_{\rm reg}^2+{\cal C}_3M_{\rm reg}\right)\ln\frac{m_n^2}{M_{\rm reg}^2} \nonumber\\
 &\quad
 +{\cal C}_4m_n^4+{\cal C}_5m_n^2M_{\rm reg}^2+{\rm \cal C}_6M_{\rm reg}^4+{\cal O}\left(\frac{M_{\rm reg}^6}{m_n^2}\right), 
\end{align}
where the coefficients~${\cal C}_i$ ($i=1,2,\cdots,6$) are functions of $\alpha$, $\beta_1$ and $\beta_2$. 
The requirement~\eqref{fourth_constraint} indicates that all ${\cal C}_i$ ($i=1,2,\cdots,6$) vanish. 
We find that ${\cal C}_1$, ${\cal C}_2$ and ${\cal C}_3$ automatically vanish, and do not give any constraints on $\alpha$, $\beta_1$ 
and $\beta_2$. 
The coefficient~${\cal C}_4$ is a function of only $\alpha$, 
\begin{align}
 {\cal C}_4 &= -\alpha^2\ln\alpha-\frac{(\alpha+2)^2}{3}\ln\frac{\alpha+2}{3}+\frac{(2\alpha+1)^2}{3}\ln\frac{2\alpha+1}{3}, 
\end{align}
and the solution of ${\cal C}_4=0$ is $\alpha=1$.  
Under the condition~$\alpha=1$, we can easily see that both ${\cal C}_5$ and ${\cal C}_6$ vanish identically.  
Therefore, we can take $\beta_2=0$, and assume that
\begin{align}
 M_1^2 &= m_n^2+\beta_1M_{\rm reg}^2, 
\end{align}
as a solution to \eqref{fourth_constraint}. 
If we rescale $M_{\rm reg}^2$, we can always set $\beta_1=1$. 
As a result, we can choose a solution of \eqref{fourth_constraint} (and \eqref{c:constraints}) as
\begin{align}\label{example:M_i}
 M_1^2(m_n^2,M_{\rm reg}^2) &= M_{\rm reg}^2+m_n^2, \nonumber\\
 M_2^2(m_n^2,M_{\rm reg}^2) &= \frac{1}{3}M_{\rm reg}^2+m_n^2, \nonumber\\
 M_3^2(m_n^2,M_{\rm reg}^2) &= \frac{2}{3}M_{\rm reg}^2+m_n^2. 
\end{align}
This result indicates that the regulators can be regarded as the KK modes for 5D bulk fields. 
In fact, if we introduce one fermionic 5D field with the (squared) bulk mass~$M_{\rm bulk}^2+M_{\rm reg}^2$, 
three fermionic 5D fields with $M_{\rm bulk}^2+\frac{1}{3}M_{\rm reg}^2$, 
and three bosonic 5D fields with $M_{\rm bulk}^2+\frac{2}{3}M_{\rm reg}^2$, 
the conditions~\eqref{c:constraints} and \eqref{lim:rhop} are satisfied for each KK mode.

Before ending this section, we comment on the relation to analytic regularization. 
In that regularization, $\rho_n$ and $p_n$ are read off from \eqref{limit_regularized_expressions} as
\begin{align}
 \rho_n &= -p_n = -\frac{\mu^4\Gamma(-\frac{d+1}{2})}{2(4\pi)^{\frac{d+1}{2}}}\left(\frac{m_n}{\mu}\right)^{d+1} 
 = -\frac{m_n^4}{32\pi^2}\left(\frac{m_n^2}{4\pi\mu^2}\right)^{\frac{d-3}{2}}\Gamma\left(\frac{3-d}{2}-2\right) \nonumber\\
 &= -\frac{m_n^4}{32\pi^2}\left\{\frac{1}{3-d}-\frac{1}{2}\left(\ln\frac{m_n^2}{4\pi\mu^2}-\frac{3}{2}+\gamma_{\rm E}\right)
 +{\cal O}\left(\frac{d-3}{2}\right)\right\}, 
\end{align}
where $\gamma_{\rm E}$ is the Euler-Mascheroni constant. 
We have used \eqref{Gamma_pole} at the last equality. 
After the minimal subtraction, we have
\begin{align}\label{result:dim_reg}
 \rho_n &= -p_n = \frac{m_n^4}{64\pi^2}\ln\frac{m_n^2}{\mu^2}. 
\end{align}
To match \eqref{PVregulated_rhop} with this result, a further additional condition has to be imposed~\cite{Koksma:2011cq}. 
\begin{align}\label{c:constraint:2}
 \sum_{i=1}^kc_iM_i^4\ln\frac{M_i^2}{\mu^2} &= 0. 
\end{align}
Therefore, the number of the regulator species has to be chosen as $k\geq 4$. 
Then, \eqref{PVregulated_rhop} agrees with \eqref{result:dim_reg} 
in the limit of $\Lambda_{\rm cut}\to\infty$. 
However, we do not have a simple solution of \eqref{c:constraints} when $k\geq 4$.  
For example, if we assume that $k=4$ and $c_3=-c_4$, we obtain from (\ref{c:constraints}) 
\begin{align}
 M_3^2 &= -{\cal A}+{\cal B}, \nonumber\\
 M_4^2 &= {\cal A}+{\cal B},  
\end{align}
where
\begin{align}
 {\cal A} &\equiv \frac{c_1M_1^2+(1-c_1)M_2^2-m_n^2}{2c_3}, \nonumber\\
 {\cal B} &\equiv \frac{c_1M_1^4+(1-c_1)M_2^4}{4c_3{\cal A}}. 
\end{align}
Plugging this into \eqref{c:constraint:2} and solving it, we can express $M_2^2$ in terms of $M_1^2$ in principle. 
As a result, $M_i^2$ ($i=2,3,4$) can be expressed as functions of $M_1^2$ and $m_n^2$. 
However, we do not have analytic expressions for them in general. 

As we will see in the next section, 
even if the condition~\eqref{c:constraint:2} is not imposed, 
the result well agrees with the one obtained 
in the analytic regularization~\eqref{eq:rho-z-regularization2} as long as ${\rm min}\,(M_1^2,M_2^2,M_3^2)>m_{\rm KK}\equiv R^{-1}$,

\section{Regulator-mass dependence of Casimir energy}
\label{sec:UV-dependence}
\begin{figure}[t]
\begin{center}
 \includegraphics[scale=0.65]{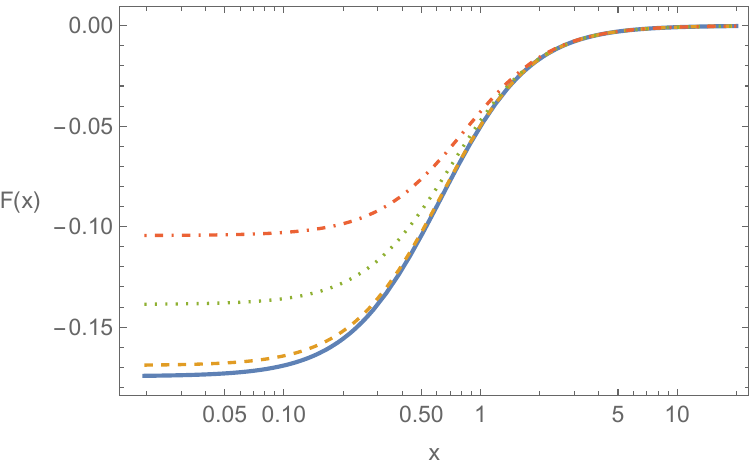}
\end{center}
 \caption{The profile of the function~$F(x)$ defined in \eqref{expr:MsF}.  
 The bulk mass is chosen as $\hat{M}_{\rm bulk}=0$ (solid), 0.1 (dashed), 0.3 (dotted) and 0.5 (dotdashed) from bottom to top. }
\label{fig:profileF}
\end{figure}

In this section, we will numerically calculate the Casimir energy density and pressure 
in the Pauli-Villars regularization, and evaluate their dependence on the regulator mass scale~$M_{\rm reg}$. 
As a specific example, we choose the regulator masses as \eqref{example:M_i}. 
In this case, the energy density and pressure for the vacuum are expressed as
\begin{align}\label{expr:rho^PV}
 \Braket{0|\hat{\rho}|0}^{\rm PV} &= -\Braket{0|\hat{p}|0}^{\rm PV} \nonumber\\
 &= \frac{M_{\rm reg}^4}{64\pi^2}\sum_{n=1}^\infty\left(\hat{m}_n^4\ln\hat{m}_n^2
 -\sum_{i=1}^3c_i\hat{M}_i^4\ln\hat{M}_i^2\right) 
 \nonumber\\
 &= \frac{M_{\rm reg}^4}{64\pi^2}\sum_{n=1}^\infty F(an), 
\end{align}
where $a\equiv (M_{\rm reg}R)^{-1}$, and 
\begin{align}\label{expr:MsF}
 \hat{m}_n^2 &\equiv \frac{m_n^2}{M_{\rm reg}^2} = \hat{M}_{\rm bulk}^2+a^2n^2,  \;\;\;\;\;
  \hat{M}_{\rm bulk} \equiv \frac{M_{\rm bulk}}{M_{\rm reg}},  \nonumber\\
 \hat{M}_1^2 &\equiv \frac{M_1^2}{M_{\rm reg}^2} = 1+\frac{m_n^2}{M_{\rm reg}^2} 
 = 1+\left(\hat{M}_{\rm bulk}^2+a^2n^2\right), \nonumber\\
 \hat{M}_2^2 &\equiv \frac{M_2^2}{M_{\rm reg}^2} 
 = \frac{1}{3}+\frac{m_n^2}{M_{\rm reg}^2} = \frac{1}{3}+\left(\hat{M}_{\rm bulk}^2+a^2n^2\right), \nonumber\\
 \hat{M}_3^2 &\equiv \frac{M_3^2}{M_{\rm reg}^2} = \frac{2}{3}+\frac{m_n^2}{M_{\rm reg}^2} 
 = \frac{2}{3}+\left(\hat{M}_{\rm bulk}^2+a^2n^2\right), \nonumber\\
 F(x) 
 &\equiv \left(\hat{M}_{\rm bulk}^2+x^2\right)^2\ln\left(\hat{M}_{\rm bulk}^2+x^2\right) \nonumber\\
 &\quad
 -\left(1+\hat{M}_{\rm bulk}^2+x^2\right)^2\ln\left(1+\hat{M}_{\rm bulk}^2+x^2\right) \nonumber\\
 &\quad
 -3\left(\frac{1}{3}+\hat{M}_{\rm bulk}^2+x^2\right)^2\ln\left(\frac{1}{3}+\hat{M}_{\rm bulk}^2+x^2\right) \nonumber\\
 &\quad
 +3\left(\frac{2}{3}+\hat{M}_{\rm bulk}^2+x^2\right)^2\ln\left(\frac{2}{3}+\hat{M}_{\rm bulk}^2+x^2\right). 
\end{align}
Fig.~\ref{fig:profileF} shows the profile of the function~$F(x)$ for various values of $\hat{M}_{\rm bulk}$. 
We can see that the contribution of the KK modes damps around $x=1$, which corresponds to the regulator mass scale~$M_{\rm reg}$.

According to \eqref{def:rho_Casimir}, the Casimir energy and pressure are given by 
\begin{align}\label{E_Casimir^PV}
 \Braket{0|\hat{\rho}|0}_{\rm Casimir}^{\rm PV} &= -\Braket{0|\hat{p}|0}_{\rm Casimir}^{\rm PV} 
 = \frac{M_{\rm reg}^4}{64\pi^2}\Delta(a),  
\end{align}
where~\footnote{
We have used that
\begin{align}
 \lim_{a\to 0}\sum_{n=1}^\infty aF(an) = \int_0^\infty dx\;F(x)-\frac{a}{2}F(0) 
 = a\int_0^\infty dx\;F(ax)-\frac{a}{2}F(0). 
\end{align}
See Sec.~3.3 of Ref.~\cite{Asai:2021csl} for details.
}
\begin{align}\label{def:Delta}
 \Delta(a) &\equiv\sum_{n=1}^\infty F(an)-\int_0^\infty dx\;F(ax)+\frac{1}{2}F(0). 
\end{align}

In order to evaluate~$\Delta(a)$, the Euler-Maclaurin formula is useful~\cite{Boyer:1968uf,Mahajan:2006mw,Saghian:2012zy}. 
Then we obtain
\begin{align}\label{expr:Delta}
 \Delta(a) &= \lim_{N_{\rm cut}\to\infty}\left\{\sum_{n=1}^{N_{\rm cut}}F(an)
 -\int_0^{N_{\rm cut}} dx\;F(ax)+\frac{1}{2}F(0)\right\} \nonumber\\
 &= \lim_{N_{\rm cut}\to\infty}\left\{\sum_{n=0}^{N_{\rm cut}}F(an)-\int_0^{N_{\rm cut}}dx\;F(ax)\right\}
 -\frac{1}{2}F(0) \nonumber\\
 &= \lim_{N_{\rm cut}\to\infty}\bigg[\frac{1}{2}\left\{F(0)+F(aN_{\rm cut})\right\} \nonumber\\
 &\hspace{20mm}\left.
 +\sum_{p=1}^{{\rm floor}\,(q/2)}\frac{B_{2p}a^{2p-1}}{(2p)!}\left\{F^{(2p-1)}(aN_{\rm cut})-F^{(2p-1)}(0)\right\}+R_q\right]-\frac{F(0)}{2}  
 \nonumber\\
 &= -\sum_{p=1}^{{\rm floor}\,(q/2)}\frac{B_{2p}a^{2p-1}}{(2p)!}F^{(2p-1)}(0)+\lim_{N_{\rm cut}\to\infty}R_q, 
\end{align}
where $B_{2p}$ are the Bernoulli numbers, $q$ is an integer greater than 1, and 
\begin{align}
 R_q &\equiv (-1)^{q-1}\int_0^{N_{\rm cut}} dx\;\frac{B_q(x-{\rm floor}\,(x))}{q!}a^qF^{(q)}(ax), 
\end{align}
with the Bernoulli polynomial~$B_q(x)$. 
At the last step in \eqref{expr:Delta}, we have used that 
\begin{align}
 \lim_{x\to\infty}F(x) &= \lim_{x\to\infty}F^{(1)}(x) = \cdots = \lim_{x\to\infty}F^{(q-1)}(x)=0. 
\end{align}
Here we set $q=2$. 
Then, noting that $F^{(1)}(0)=0$ from \eqref{derivatives:F}, \eqref{expr:Delta} becomes
\begin{align}
 \Delta(a) &= -\int_0^\infty dx\;\frac{B_2(x-{\rm floor}(x))a^2}{2}F^{(2)}(ax) \nonumber\\
 &= -\frac{a^2}{2}\sum_{l=0}^\infty \int_0^1 dx\;B_2(x)F^{(2)}(a(x+l)), 
\end{align}
where the explicit form of $F^{(2)}(x)$ is shown in \eqref{derivatives:F} in Appendix, and
\begin{align}
 B_2(x) &= x^2-x+\frac{1}{6}. 
\end{align}

\begin{figure}[t]
\begin{center}
 \includegraphics[scale=0.65]{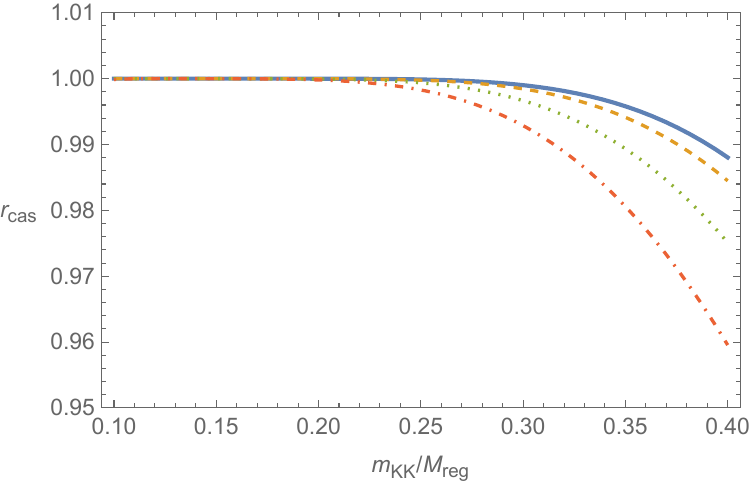}
\end{center}
 \caption{The ratio~$r_{\rm cas}$ defined in \eqref{def:r_cas} as a function of $m_{\rm KK}/M_{\rm reg}$. 
 The bulk mass is chosen as $M_{\rm bulk}/M_{\rm reg}=0$ (solid), 0.1 (dashed), 0.2 (dotted), and 0.3 (dotdashed), respectively. }
\label{fig:r_cas}
\end{figure}

To see the deviation of the Casimir energy~\eqref{E_Casimir^PV} 
from the one obtained in the analytic regularization~\eqref{eq:rho-z-regularization2}, 
we define 
\begin{align}\label{def:r_cas}
 r_{\rm cas} &\equiv \frac{\Braket{0|\hat{\rho}|0}_{\rm Casimir}^{\rm PV}}{\Braket{0|\hat{\rho}|0}_{\rm Casimir}^{\rm anal}}, 
\end{align}
where $\Braket{0|\hat{\rho}|0}_{\rm Casimir}^{\rm anal}$ denotes \eqref{eq:rho-z-regularization2}. 
Fig.~\ref{fig:r_cas} shows the ratio~$r_{\rm cas}$ as a function of $a=m_{\rm KK}/M_{\rm reg}$, 
where $m_{KK}\equiv 1/R$ is the KK mass scale. 
We can see that the result obtained by the Pauli-Villars regularization well agrees with that of the analytic regularization 
as long as the compactification scale~$m_{\rm KK}$ is well below the regulator mass scale~$M_{\rm reg}$. 

Before ending the section, one comment is in order. 
The above results can also be expressed by using the analytic regularized formula~\eqref{eq:rho-z-regularization2}. 
As mentioned below \eqref{example:M_i}, 
the current choice of the Pauli-Villars regulators can be understood as 5D fields. 
Thus, the Casimir energy density in \eqref{expr:rho^PV} is also expressed as 
\begin{align}
 \langle 0|\hat{\rho}|0\rangle_{\rm Casimir}^{\rm PV} &= {\cal E}(R,M_{\rm bulk})-{\cal E}\left(R,\sqrt{M_{\rm bulk}^2+M_{\rm reg}^2}\right) \nonumber\\
 &\quad
 -3{\cal E}\left(R,\sqrt{M_{\rm bulk}^2+\frac{1}{3}M_{\rm reg}^2}\right)
 +3{\cal E}\left(R,\sqrt{M_{\rm bulk}^2+\frac{2}{3}M_{\rm reg}^2}\right), 
\end{align}
where
\begin{align}\label{def:cE}
 {\cal E}(R,M) &\equiv -\frac{M^{\frac{5}{2}}}{16\pi^4R^{\frac{3}{2}}}
 \sum_{n=1}^\infty n^{-\frac{5}{2}}K_{\frac{5}{2}}(2\pi nRM). 
\end{align}
Thus, \eqref{def:r_cas} can be rewritten as
\begin{align}
 r_{\rm cas} &= 1-\Delta\left(\bar{M}_{\rm bulk},\sqrt{\bar{M}_{\rm bulk}^2+\bar{M}_{\rm reg}^2}\right) \nonumber\\
 &\quad
 -3\Delta\left(\bar{M}_{\rm bulk},\sqrt{\bar{M}_{\rm bulk}^2+\frac{1}{3}\bar{M}_{\rm reg}^2}\right)
 +3\Delta\left(\bar{M}_{\rm bulk},\sqrt{\bar{M}_{\rm bulk}^2+\frac{2}{3}\bar{M}_{\rm reg}^2}\right), 
\end{align}
where $\bar{M}_{\rm bulk}=RM_{\rm bulk}$, $\bar{M}_{\rm reg}=RM_{\rm reg}$, and 
\begin{align}\label{def:Dlt}
 \Delta(\bar{M}_1,\bar{M}_2) &\equiv \left(\frac{\bar{M}_2}{\bar{M}_1}\right)^{\frac{5}{2}}
 \frac{\sum_{n=1}^\infty n^{-\frac{5}{2}}K_{\frac{5}{2}}(2\pi n\bar{M}_2)}{\sum_{n=1}^\infty n^{-\frac{5}{2}}K_{\frac{5}{2}}(2\pi n\bar{M}_1)}. 
\end{align}
The function $\Delta(\bar{M}_1,\bar{M}_2)$ is exponentially suppressed when $\bar{M}_1\ll\bar{M}_2$, 
but becomes non-negligible when $\bar{M}_2={\cal O}(\bar{M}_1)$. 
Since the infinite summation in \eqref{def:cE} or \eqref{def:Dlt} converges much faster than the KK summation, this expression is convenient to 
the numerical computation.

\section{Discussions and Conclusions}
\label{sec:conclusions}

We studied the dependence of the Casimir energy density on the UV dynamics 
in the context of a 5D model with a compact dimension. 
In contrast to renormalizable theories, a non-renormalizable theory, such as our 5D model, should be regarded as
an effective theory, and be replaced by a more fundamental theory at some high energy scale~$M_{\rm UV}$. 
A typical situation is that some new particles appear at a scale around $M_{\rm UV}$, 
and cancel quantum corrections from the light fields in the 5D effective theory. 

If $M_{\rm UV}$ is not far from $m_{\rm KK}$, the existence of the new particles can affect low-energy observables, 
such as the Casimir energy density. 
We have evaluated such effects on the Casimir energy density (and pressure). 
The most popular way of calculating the Casimir energy is the method using the analytic regularization 
because the resultant expression is convenient for the numerical evaluation and the regularization preserves 
various symmetries, including the Lorentz symmetry. 
However, this regularization removes the power-law divergences by hand, 
and thus is inappropriate for our purpose, as we showed in Sec.~\ref{UVsensitivity_analytic}. 
Instead of this, we work in the Pauli-Villars regularization, which mimics the situation that new particles 
cancel the quantum corrections from the light particles. 
To preserve the Lorentz symmetry of the vacuum, we have to prepare more than one regulator for each mode, 
and their masses and the degrees of freedom have to satisfy some conditions (see \eqref{c:constraints} and \eqref{lim:rhop}). 
It should be noticed that two of them are automatically satisfied in a (spontaneously broken) SUSY model. 
The result in \eqref{example:M_i} indicates that the Pauli-Villars regulators can be regarded as the KK modes for 5D bulk fields. 
In a case that the model is embedded into a (spontaneously broken) SUSY 5D theory, 
the scalar field~$\Phi$ and the bulk regulators should be embedded into a 5D SUSY multiplet. 
Thus, the example of the regulators considered in Sec.~\ref{sec:Pauli-Villars} must be modified. 
Needless to say, the deviation from the result in the analytic regularization depends on the choice of the Pauli-Villars regulators. 
Still, our example shows a typical order of magnitude for the deviation. 

If we do not impose the condition~\eqref{c:constraint:2}, 
it is not guaranteed that the resultant Casimir energy density (or pressure) agrees with 
the one obtained by the analytic regularization. 
We numerically evaluate them and confirm that they well agree with each other 
even if \eqref{c:constraint:2} is not satisfied, 
as loong as the KK mass~$m_{\rm KK}$ and the bulk mass~$M_{\rm bulk}$ are smaller than all the regulator masses.

\section*{Acknowledgment}
H.M. would like to thank Yoshio Matsumoto for useful comments and for checking our manuscripts in the early stages. The work of H.M. was supported by JSPS KAKENHI Grant No. JP22KJ1782 and No. JP23K13100.

\appendix

\section{Complete and incomplete beta and gamma functions}
\label{appendix:i-beta-function}

\subsection{Definitions and properties}

The integral expressions of the complete beta and gamma functions are given by
\begin{eqnarray}
B(\alpha, \beta) &\equiv& \int_0^\infty dx\; x^{\alpha-1}(1-x)^{\beta-1} = B(\beta, \alpha), \nonumber\\
\Gamma(\alpha) &\equiv& \int_0^\infty dt\; t^{\alpha-1}e^{-t}, 
\end{eqnarray}
which are valid only for ${\rm Re}\,\alpha>0$ and ${\rm Re}\,\beta>0$.
They are related as
\begin{equation}
B(\alpha, \beta) = \frac{\Gamma(\alpha)\Gamma(\beta)}{\Gamma(\alpha+\beta)}. \label{rel:bt-gm}
\end{equation}
This relation holds over the whole domain of the beta function.
The gamma function behaves near $\alpha=0,-1,-2$ as
\begin{align}\label{Gamma_pole}
 \Gamma(\epsilon) &= \frac{1}{\epsilon}-\gamma_{\rm E}+{\cal O}(\epsilon), \nonumber\\
 \Gamma(-1+\epsilon) &= -\frac{1}{\epsilon}-1+\gamma_{\rm E}+{\cal O}(\epsilon), \nonumber\\
 \Gamma(-2+\epsilon) &= \frac{1}{2\epsilon}+\frac{3}{4}-\frac{\gamma_{\rm E}}{2}+{\cal O}(\epsilon), 
\end{align}
where $\gamma_{\rm E}$ is the Euler-Mascheroni constant. 

The incomplete beta functions are defined as
\begin{eqnarray}
B_z(\alpha, \beta) &\equiv& \int_0^z dx\; x^{\alpha-1}(1-x)^{\beta-1}, 
\label{def:bt:incomplete}
\end{eqnarray}
for ${\rm Re}\,\alpha>0$, 
and the upper and the lower incomplete gamma functions are defined as
\begin{eqnarray}
\Gamma_z(\alpha) &\equiv& \int_z^\infty dt\; t^{\alpha-1}e^{-t}, \nonumber\\
\gamma_z(\alpha) &\equiv& \Gamma(\alpha)-\Gamma_z(\alpha) = \int_0^z dt\; t^{\alpha-1}e^{-t}.
\end{eqnarray}
where the integral expression of $\gamma_z(\alpha)$ is valid only for ${\rm Re}\,\alpha>0$. 
From \eqref{def:bt:incomplete}, we obtain
\begin{align}\label{expand:incomplete_beta}
 B_{1-\epsilon}(\alpha,\beta) &= \int_\epsilon^1 dy\;y^{\beta-1}(1-y)^{\alpha-1} \;\;\;\;\; (y\equiv 1-x) \nonumber\\
 &= \int_0^1dy\;y^{\beta-1}(1-y)^{\alpha-1}-\int_0^\epsilon dy\;y^{\beta-1}(1-y)^{\alpha-1} \nonumber\\
 &= B(\beta,\alpha)-\int_0^\epsilon dy\;y^{\beta-1}\left\{1-(\alpha-1)y+\frac{2-3\alpha+\alpha^2}{2}y^2+{\cal O}(y^3)\right\} \nonumber\\
 &= B(\alpha,\beta)-\frac{\epsilon^\beta}{\beta}+\frac{\alpha-1}{\beta+1}\epsilon^{\beta+1}
 -\frac{\alpha^2-3\alpha+2}{2(\beta+2)}\epsilon^{\beta+2}+{\cal O}(\epsilon^{\beta+3}), 
\end{align}
for ${\rm Re}\,\alpha>0$ and ${\rm Re}\,\beta>0$. 

Similarly, the incomplete gamma function can be expanded as
\begin{align}\label{expand:incompleteGamma}
 \Gamma_\delta(\alpha) &= \int_0^\infty dt\;t^{\alpha-1}e^{-t}-\int_0^\delta dt\;t^{\alpha-1}e^{-t} \nonumber\\
 &= \Gamma(\alpha)-\int_0^\delta dt\;t^{\alpha-1}\left\{1-t+\frac{t^2}{2}+{\cal O}(t^3)\right\} \nonumber\\
 &= \Gamma(\alpha)-\frac{\delta^\alpha}{\alpha}+\frac{\delta^{\alpha+1}}{\alpha+1}
 -\frac{\delta^{\alpha+2}}{2(\alpha+2)}+{\cal O}(\delta^{\alpha+3}). 
\end{align}

The incomplete beta function is also expanded as
\begin{align}\label{integral:incompleteBeta}
 B_{1-\epsilon}(\alpha,\beta) &= \frac{1}{\alpha\Gamma(\alpha+\beta)}\int_0^\infty dx\;x^\alpha e^{-x}
 \Gamma_{\frac{x\epsilon}{1-\epsilon}}(\beta)+\frac{1}{\alpha}\epsilon^{\beta}(1-\epsilon)^{\alpha}. 
\end{align}
We can show this by differentiating both hand sides concerning $\epsilon$, 
and checking that they coincide. 
For $\beta>0$, \eqref{integral:incompleteBeta} reduces to \eqref{rel:bt-gm} in the limit of $\epsilon\to 0$.

\subsection{Explicit forms}

Here we show the explicit forms of the incomplete beta functions that appear in Sec.~\ref{cutoff_reg}, 
\begin{align}
 B_{1-\epsilon_n}\left(\frac{3}{2},-2\right), \;\;\;\;\;
 B_{1-\epsilon_n}\left(\frac{5}{2},-2\right). 
\end{align}
Since 
\begin{align}
 1-\epsilon_n &= 1-\frac{m_n^2}{\Lambda_{\rm cut}^2+m_n^2} = \frac{\Lambda_{\rm cut}^2}{\Lambda_{\rm cut}^2+m_n^2}
 = \frac{X^2}{X^2+1}, 
\end{align}
where $X\equiv \Lambda_{\rm cut}/m_n$, the above functions can be expressed in the form of 
\begin{align}\label{expr:icBeta}
 B_{\frac{X^2}{X^2+1}}(\alpha,\beta) &= \int_0^{\frac{X^2}{X^2+1}} dx\;x^{\alpha-1}(1-x)^{\beta-1}, 
\end{align}
when ${\rm Re}\,\alpha>0$. 
By differentiating this concerning $X$, we have
\begin{align}
 \partial_X B_{\frac{X^2}{X^2+1}}(\alpha,\beta) &= \left(\frac{X^2}{X^2+1}\right)^{\alpha-1}\left(1-\frac{X^2}{X^2+1}\right)^{\beta-1}
 \partial_X\left(\frac{X^2}{X^2+1}\right) \nonumber\\
 &= \frac{2X^{2\alpha-1}}{(X^2+1)^{\alpha+\beta}}. 
\end{align}
Since $B_{\frac{X^2}{X^2+1}}(\alpha,\beta)|_{X=0}=0$, \eqref{expr:icBeta} is reexpressed as
\begin{align}
 B_{\frac{X^2}{X^2+1}}(\alpha,\beta) &= \int_0^XdY\;\frac{2Y^{2\alpha-1}}{(Y^2+1)^{\alpha+\beta}}. 
\end{align}
From this expression, we obtain
\begin{align}\label{explicit_icBs}
 B_{\frac{X^2}{X^2+1}}\left(\frac{3}{2},-2\right) &= \int_0^X dY\;2Y^2\sqrt{Y^2+1} \nonumber\\
 &= \frac{1}{4}\left\{X\sqrt{X^2+1}(2X^2+1)-\ln\left(X+\sqrt{X^2+1}\right)\right\}, \nonumber\\
 B_{\frac{X^2}{X^2+1}}\left(\frac{5}{2},-2\right) &= \int_0^X dY\;\frac{2Y^4}{\sqrt{Y^2+1}} \nonumber\\
 &= \frac{1}{4}\left\{X\sqrt{X^2+1}(2X^2-3)+3\ln\left(X+\sqrt{X^2+1}\right)\right\}. 
\end{align}

\section{Formulae for zeta function regularization}
\label{appendix:z-regularization}

In order to evaluate the regularized sums in \eqref{regularized_expressions}, we define
\begin{align}\label{def:U}
 U(\alpha,\beta;M^2) &\equiv \sum_{n=1}^\infty B_{1-\epsilon_n}(\alpha,\beta)\left(M^2+n^2\right)^{-\beta}e^{-a^2n^2}, 
\end{align}
where $a$ is a tiny positive constant, and
\begin{align}
 \epsilon_n &\equiv \frac{M^2+n^2}{a^{-2}+M^2+n^2}. 
\end{align}
Using the formula~\eqref{integral:incompleteBeta}, this is expressed as
\begin{align}\label{expr:U}
 U(\alpha,\beta;M^2) &= \sum_{n=1}^\infty\left\{
 \int_0^\infty dx\;\frac{x^\alpha e^{-x}}{\alpha\Gamma(\alpha+\beta)}\Gamma_{(M^2+n^2)a^2x}(\beta)
 +\frac{\epsilon_n^\beta(1-\epsilon_n)^\alpha}{\alpha}\right\} \nonumber\\
 &\quad
 \times \left(M^2+n^2\right)^{-\beta}e^{-a^2n^2} \nonumber\\
 &= \frac{U_1(\alpha,\beta;M^2)}{\alpha(\alpha+\beta)}
 +\frac{U_2(\alpha,\beta;M^2)}{\alpha a^{2\alpha}}, 
\end{align}
where
\begin{align}\label{def:Us}
 U_1(\alpha,\beta;M^2) &\equiv \int_0^\infty dx\;x^\alpha e^{-x}S_{a^2x}(\beta;M^2), \nonumber\\
 S_\delta(\beta;M^2) &\equiv \sum_{n=1}^\infty \left(M^2+n^2\right)^{-\beta}\Gamma_{(M^2+n^2)\delta}(\beta)e^{-a^2n^2}, 
 \nonumber\\
 U_2(\alpha,\beta;M^2) &\equiv \sum_{n=1}^\infty \frac{e^{-a^2n^2}}{\left(a^{-2}+M^2+n^2\right)^{\alpha+\beta}}. 
\end{align}
Here note that $S_\delta(\beta;M^2)$ can be rewritten as
\begin{align}
 S_\delta(\beta;M^2) &= \int_\delta^\infty dt\;t^{\beta-1}e^{-M^2t}\vartheta(t+a^2), 
\end{align}
where $\vartheta(t)\equiv \sum_{n=1}^\infty e^{-n^2t}$ is the Jacobi theta function, which has the property, 
\begin{align}
 \vartheta(t) &= -\frac{1}{2}+\frac{1}{2}\sqrt{\frac{\pi}{t}}+\sqrt{\frac{\pi}{t}}\vartheta\left(\frac{\pi^2}{t}\right). 
\end{align}
Using this property of $\vartheta(t)$ and the definition of the gamma function, $S_{a^2x}(\beta;M^2)$ is expanded as
\begin{align}\label{expr:S_dlt}
 S_{a^2x}(\beta;M^2) &= -\frac{M^{-2\beta}}{2}\Gamma(\beta)
 +\frac{\sqrt{\pi}M^{-2\beta+1}}{2}\sum_{j=0}^\infty c_ja^{2j}\Gamma\left(\beta-\frac{1}{2}-j\right) \nonumber\\
 &\quad
 +\frac{(a^2x)^\beta}{2\beta}-\frac{M^2(a^2x)^{\beta+1}}{2(\beta+1)}+\frac{M^4(a^2x)^{\beta+2}}{4(\beta+2)} \nonumber\\
 &\quad
 -\frac{\sqrt{\pi}M^{-2\beta+1}}{2}\left\{
 a^{2\beta-1}H_{\beta-\frac{1}{2}}(M^2x)
 -a^{2\beta+1}H_{\beta+\frac{1}{2}}(M^2x) \right.\nonumber\\
 &\hspace{30mm}\left.
 +\frac{a^{2\beta+3}}{2}H_{\beta+\frac{3}{2}}(M^2x)\right\} \nonumber\\
 &\quad
 +\sqrt{\pi}\int_\delta^\infty dt\;\frac{t^{\beta-1}e^{-M^2t}}{\sqrt{t+a^2}}\vartheta\left(\frac{\pi^2}{t+a^2}\right)
 +{\cal O}\left(a^{2\beta+5}\right), 
\end{align}
where the constants~$c_j$ are defined by
\begin{align}
 \left(1+x\right)^{-1/2} &= \sum_{j=0}^\infty c_jx^j, 
\end{align}
and the function~$H_b(z)$ is defined as 
\begin{align}
 H_b(z) &\equiv \sum_{j=0}^\infty\frac{c_j}{b-j}z^{b-j}. 
\end{align}
When ${\rm Re}\,\beta>\frac{1}{2}$, all the powers of $a$ are positive, and we can take the limit of $a\to 0$ and obtain 
\begin{align}\label{def:S_0}
 S_0(\beta;M^2) &\equiv \lim_{a\to 0}S_{a^2x}(\beta;M^2) = \Gamma(\beta)\sum_{n=1}^\infty (M^2+n^2)^{-\beta} \nonumber\\
 &= -\frac{M^{-2\beta}}{2}\Gamma(\beta)+\frac{\sqrt{\pi}M^{-2\beta+1}}{2}\Gamma\left(\beta-\frac{1}{2}\right) \nonumber\\
 &\quad
 +\sqrt{\pi}\sum_{n=1}^\infty\int_0^\infty dt\;t^{\beta-\frac{3}{2}}\exp\left(-M^2t-\frac{\pi^2 n^2}{t}\right). 
\end{align}
When ${\rm Re}\,\beta<1$, the integral in the last term is expressed as
\begin{align}\label{int-K}
 \int_0^\infty dt\;t^{\beta-\frac{3}{2}}\exp\left(-M^2t-\frac{\pi^2 n^2}{t}\right) 
 &= 2\left(\frac{\pi n}{M}\right)^{\beta-\frac{1}{2}}K_{\frac{1}{2}-\beta}(2\pi nM). 
\end{align}

Substituting \eqref{expr:S_dlt} into the first expression in \eqref{def:Us}, we obtain
\begin{align}\label{expr:U1}
 U_1(\alpha,\beta;M^2) &= -\frac{M^{-2\beta}\alpha\Gamma(\alpha)\Gamma(\beta)}{2}
 +\frac{\sqrt{\pi}M^{-2\beta+1}\Gamma(\alpha)}{2}\sum_{j=0}^\infty c_ja^{2j}\left(\beta-\frac{1}{2}-j\right) \nonumber\\
 &\quad
 +\frac{\Gamma(\alpha+\beta+1)}{2\beta}a^{2\beta}
 -\frac{M^2\Gamma(\alpha+\beta+2)}{2(\beta+1)}a^{2\beta+2}
 +\frac{M^4\Gamma(\alpha+\beta+3)}{4(\beta+2)}a^{2\beta+4} \nonumber\\
 &\quad
 +C_1(\beta;M^2)a^{2\beta-1}+C_2(\beta;M^2)a^{2\beta+1}+C_3(\beta;M^2)a^{2\beta+3} \nonumber\\
 &\quad
 +\sqrt{\pi}\int_0^\infty dx\;x^\alpha e^{-x}\int_{a^2x}^\infty dt\;\frac{t^{\beta-1}e^{-M^2t}}{\sqrt{t+a^2}}
 \vartheta\left(\frac{\pi^2}{t+a^2}\right)+{\cal O}\left(a^{2\beta+5}\right), 
\end{align}
where
\begin{align}\label{def:Cs}
 C_1(\beta;M^2) &\equiv -\frac{\sqrt{\pi}M^{-2\alpha-2\beta}}{2}
 \int_0^\infty dy\;y^\alpha e^{-y/M^2}H_{\beta-\frac{1}{2}}(y), \nonumber\\
 C_2(\beta;M^2) &\equiv \frac{\sqrt{\pi}M^{-2\alpha-2\beta}}{2}
 \int_0^\infty dy\;y^\alpha e^{-y/M^2}H_{\beta+\frac{1}{2}}(y), \nonumber\\
 C_3(\beta;M^2) &\equiv -\frac{\sqrt{\pi}M^{-2\alpha-2\beta}}{4}
 \int_0^\infty dy\;y^\alpha e^{-y/M^2}H_{\beta+\frac{3}{2}}(y). 
\end{align}

\section{Derivatives of $\mbox{\boldmath $F(x)$}$}
\label{explicit_form:cC_i}
Here we collect the explicit forms of derivatives of $F(x)$ defined in (\ref{expr:MsF}). 
\begin{align}\label{derivatives:F}
 F^{(1)}(x) &= 4x\left\{\left(\hat{M}_{\rm bulk}^2+x^2\right)\ln\left(\hat{M}_{\rm bulk}^2+x^2\right) \right.\nonumber\\
 &\hspace{12mm}
 -\left(1+\hat{M}_{\rm bulk}^2+x^2\right)\ln\left(1+\hat{M}_{\rm bulk}^2+x^2\right) \nonumber\\
 &\hspace{12mm}
 -3\left(\frac{1}{3}+\hat{M}_{\rm bulk}^2+x^2\right)\ln\left(\frac{1}{3}+\hat{M}_{\rm bulk}^2+x^2\right) \nonumber\\
 &\hspace{12mm}\left.
 +3\left(\frac{2}{3}+\hat{M}_{\rm bulk}^2+x^2\right)\ln\left(\frac{2}{3}+\hat{M}_{\rm bulk}^2+x^2\right)\right\}, \nonumber\\
 F^{(2)}(x) &= 12\left(\frac{\hat{M}_{\rm bulk}^2}{3}+x^2\right)\ln\left(\hat{M}_{\rm bulk}^2+x^2\right) \nonumber\\
 &\quad
 -12\left(\frac{1+\hat{M}_{\rm bulk}^2}{3}+x^2\right)\ln\left(1+\hat{M}_{\rm bulk}^2+x^2\right) \nonumber\\
 &\quad 
 -36\left(\frac{1}{9}+\frac{\hat{M}_{\rm bulk}^2}{3}+x^2\right)\ln\left(\frac{1}{3}+\hat{M}_{\rm bulk}^2+x^2\right) \nonumber\\
 &\quad
 +36\left(\frac{2}{9}+\frac{\hat{M}_{\rm bulk}^2}{3}+x^2\right)\ln\left(\frac{2}{3}+\hat{M}_{\rm bulk}^2+x^2\right). 
\end{align}

For $x\gg 1$, $F^{(2)}(x)$ is expanded as
\begin{align}
 F^{(2)}(x) &= -\frac{4}{9x^4}+\frac{20\left(1+2\hat{M}_{\rm bulk}^2\right)}{27x^6}
 -\frac{14}{81x^8}\left(5+18\hat{M}_{\rm bulk}^2+18\hat{M}_{\rm bulk}^4\right) \nonumber\\
 &\quad
 +\frac{8}{9x^{10}}\left(1+5\hat{M}_{\rm bulk}^2+9\hat{M}_{\rm bulk}^4+6\hat{M}_{\rm bulk}^6\right) \nonumber\\
 &\quad
 -\frac{44}{2187x^{12}}\left\{43+135\hat{M}_{\rm bulk}^2\left(1+\hat{M}_{\rm bulk}^2\right)
 \left(2+3\hat{M}_{\rm bulk}^2+3\hat{M}_{\rm bulk}^4\right)\right\} \nonumber\\
 &\quad
 +{\cal O}\left(x^{-14}\right). 
\end{align}

\bibliography{Refs}

\providecommand{\href}[2]{#2}\begingroup\raggedright\begin{thebibliography}{10}

\bibitem{Casimir:1948dh}
H.B.G.~Casimir, \emph{{On the attraction between two perfectly conducting
  plates}}, {\emph{Indag. Math.} {\bfseries 10} (1948) 261}.

\bibitem{Lamoreaux:1996wh}
S.K.~Lamoreaux, \emph{{Demonstration of the Casimir force in the 0.6 to 6
  micrometers range}},
  \href{https://doi.org/10.1103/PhysRevLett.78.5}{\emph{Phys. Rev. Lett.}
  {\bfseries 78} (1997) 5}.

\bibitem{Mohideen:1998iz}
U.~Mohideen and A.~Roy, \emph{{Precision measurement of the Casimir force from
  0.1 to 0.9 micrometers}},
  \href{https://doi.org/10.1103/PhysRevLett.81.4549}{\emph{Phys. Rev. Lett.}
  {\bfseries 81} (1998) 4549}
  [\href{https://arxiv.org/abs/physics/9805038}{{\ttfamily physics/9805038}}].

\bibitem{Roy:1999dx}
A.~Roy, C.-Y.~Lin and U.~Mohideen, \emph{{Improved precision measurement of the
  casimir force}},
  \href{https://doi.org/10.1103/PhysRevD.60.111101}{\emph{Phys. Rev. D}
  {\bfseries 60} (1999) 111101}
  [\href{https://arxiv.org/abs/quant-ph/9906062}{{\ttfamily
  quant-ph/9906062}}].

\bibitem{Bimonte:2021sib}
G.~Bimonte, B.~Spreng, P.A.~Maia~Neto, G.-L.~Ingold, G.L.~Klimchitskaya,
  V.M.~Mostepanenko et~al., \emph{{Measurement of the Casimir Force between 0.2
  and 8 $\mu$m: Experimental Procedures and Comparison with Theory}},
  \href{https://doi.org/10.3390/universe7040093}{\emph{Universe} {\bfseries 7}
  (2021) 93} [\href{https://arxiv.org/abs/2104.03857}{{\ttfamily 2104.03857}}].

\bibitem{Garriga:2000jb}
J.~Garriga, O.~Pujolas and T.~Tanaka, \emph{{Radion effective potential in the
  brane world}},
  \href{https://doi.org/10.1016/S0550-3213(01)00144-4}{\emph{Nucl. Phys. B}
  {\bfseries 605} (2001) 192}
  [\href{https://arxiv.org/abs/hep-th/0004109}{{\ttfamily hep-th/0004109}}].

\bibitem{Toms:2000bh}
D.J.~Toms, \emph{{Quantized bulk fields in the Randall-Sundrum compactification
  model}}, \href{https://doi.org/10.1016/S0370-2693(00)00618-3}{\emph{Phys.
  Lett. B} {\bfseries 484} (2000) 149}
  [\href{https://arxiv.org/abs/hep-th/0005189}{{\ttfamily hep-th/0005189}}].

\bibitem{Goldberger:2000dv}
W.D.~Goldberger and I.Z.~Rothstein, \emph{{Quantum stabilization of
  compactified AdS(5)}},
  \href{https://doi.org/10.1016/S0370-2693(00)01047-9}{\emph{Phys. Lett. B}
  {\bfseries 491} (2000) 339}
  [\href{https://arxiv.org/abs/hep-th/0007065}{{\ttfamily hep-th/0007065}}].

\bibitem{Brevik:2000vt}
I.H.~Brevik, K.A.~Milton, S.~Nojiri and S.D.~Odintsov, \emph{{Quantum
  (in)stability of a brane world AdS(5) universe at nonzero temperature}},
  \href{https://doi.org/10.1016/S0550-3213(01)00026-8}{\emph{Nucl. Phys. B}
  {\bfseries 599} (2001) 305}
  [\href{https://arxiv.org/abs/hep-th/0010205}{{\ttfamily hep-th/0010205}}].

\bibitem{Beneventano:1995fh}
C.G.~Beneventano and E.M.~Santangelo, \emph{{Connection between zeta and cutoff
  regularizations of Casimir energies}},
  \href{https://doi.org/10.1142/S0217751X96001395}{\emph{Int. J. Mod. Phys. A}
  {\bfseries 11} (1996) 2871}
  [\href{https://arxiv.org/abs/hep-th/9501122}{{\ttfamily hep-th/9501122}}].

\bibitem{Moretti:1998rf}
V.~Moretti, \emph{{Local zeta function techniques versus point splitting
  procedure: A Few rigorous results}},
  \href{https://doi.org/10.1007/s002200050558}{\emph{Commun. Math. Phys.}
  {\bfseries 201} (1999) 327}
  [\href{https://arxiv.org/abs/gr-qc/9805091}{{\ttfamily gr-qc/9805091}}].

\bibitem{Hagen:2000bu}
C.R.~Hagen, \emph{{Cutoff dependence of the Casimir effect}},
  \href{https://doi.org/10.1007/s100520100593}{\emph{Eur. Phys. J. C}
  {\bfseries 19} (2001) 677}
  [\href{https://arxiv.org/abs/quant-ph/0003108}{{\ttfamily
  quant-ph/0003108}}].

\bibitem{Visser:2016ddm}
M.~Visser, \emph{{Regularization versus Renormalization: Why Are Casimir Energy
  Differences So Often Finite?}},
  \href{https://doi.org/10.3390/particles2010002}{\emph{Particles} {\bfseries
  2} (2018) 14} [\href{https://arxiv.org/abs/1601.01374}{{\ttfamily
  1601.01374}}].

\bibitem{Matsui:2018tan}
H.~Matsui and Y.~Matsumoto, \emph{{Revisiting regularization with Kaluza-Klein
  states and Casimir vacuum energy from extra dimensional spaces}},
  \href{https://doi.org/10.1103/PhysRevD.100.016010}{\emph{Phys. Rev. D}
  {\bfseries 100} (2019) 016010}
  [\href{https://arxiv.org/abs/1804.01052}{{\ttfamily 1804.01052}}].

\bibitem{Asai:2021csl}
Y.~Asai and Y.~Sakamura, \emph{{Ultraviolet sensitivity of Casimir energy}},
  \href{https://doi.org/10.1093/ptep/ptac030}{\emph{PTEP} {\bfseries 2022}
  (2022) 033B07} [\href{https://arxiv.org/abs/2112.04708}{{\ttfamily
  2112.04708}}].

\bibitem{Akhmedov:2002ts}
E.K.~Akhmedov, \emph{{Vacuum energy and relativistic invariance}},
  \href{https://arxiv.org/abs/hep-th/0204048}{{\ttfamily hep-th/0204048}}.

\bibitem{Koksma:2011cq}
J.F.~Koksma and T.~Prokopec, \emph{{The Cosmological Constant and Lorentz
  Invariance of the Vacuum State}},
  \href{https://arxiv.org/abs/1105.6296}{{\ttfamily 1105.6296}}.

\bibitem{Martin:2012bt}
J.~Martin, \emph{{Everything You Always Wanted To Know About The Cosmological
  Constant Problem (But Were Afraid To Ask)}},
  \href{https://doi.org/10.1016/j.crhy.2012.04.008}{\emph{Comptes Rendus
  Physique} {\bfseries 13} (2012) 566}
  [\href{https://arxiv.org/abs/1205.3365}{{\ttfamily 1205.3365}}].

\bibitem{Danielsson:2018qpa}
U.~Danielsson, \emph{{The quantum swampland}},
  \href{https://doi.org/10.1007/JHEP04(2019)095}{\emph{JHEP} {\bfseries 04}
  (2019) 095} [\href{https://arxiv.org/abs/1809.04512}{{\ttfamily
  1809.04512}}].

\bibitem{Escamilla:2023oce}
L.A.~Escamilla, W.~Giar\`e, E.~Di~Valentino, R.C.~Nunes and S.~Vagnozzi,
  \emph{{The state of the dark energy equation of state circa 2023}},
  \href{https://arxiv.org/abs/2307.14802}{{\ttfamily 2307.14802}}.

\bibitem{Hawking:1991nk}
S.W.~Hawking, \emph{{The Chronology protection conjecture}},
  \href{https://doi.org/10.1103/PhysRevD.46.603}{\emph{Phys. Rev. D} {\bfseries
  46} (1992) 603}.

\bibitem{Rubakov:2014jja}
V.A.~Rubakov, \emph{{The Null Energy Condition and its violation}},
  \href{https://doi.org/10.3367/UFNe.0184.201402b.0137}{\emph{Phys. Usp.}
  {\bfseries 57} (2014) 128} [\href{https://arxiv.org/abs/1401.4024}{{\ttfamily
  1401.4024}}].

\bibitem{Visser:2016mtr}
M.~Visser, \emph{{Lorentz invariance and the zero-point stress-energy tensor}},
  \href{https://doi.org/10.3390/particles1010010}{\emph{Particles} {\bfseries
  1} (2018) 138} [\href{https://arxiv.org/abs/1610.07264}{{\ttfamily
  1610.07264}}].

\bibitem{Birrell:1982ix}
N.D.~Birrell and P.C.W.~Davies, \emph{{Quantum Fields in Curved Space}},
  Cambridge Monographs on Mathematical Physics, Cambridge Univ. Press,
  Cambridge, UK (2, 1984),
  \href{https://doi.org/10.1017/CBO9780511622632}{10.1017/CBO9780511622632}.

\bibitem{Buchbinder:1992rb}
I.L.~Buchbinder, S.D.~Odintsov and I.L.~Shapiro, \emph{{Effective action in
  quantum gravity}} (1992).

\bibitem{Stelle:1976gc}
K.S.~Stelle, \emph{{Renormalization of Higher Derivative Quantum Gravity}},
  \href{https://doi.org/10.1103/PhysRevD.16.953}{\emph{Phys. Rev. D} {\bfseries
  16} (1977) 953}.

\bibitem{Horowitz:1978fq}
G.T.~Horowitz and R.M.~Wald, \emph{{Dynamics of Einstein's Equation Modified by
  a Higher Order Derivative Term}},
  \href{https://doi.org/10.1103/PhysRevD.17.414}{\emph{Phys. Rev. D} {\bfseries
  17} (1978) 414}.

\bibitem{Horowitz:1980fj}
G.T.~Horowitz, \emph{{SEMICLASSICAL RELATIVITY: THE WEAK FIELD LIMIT}},
  \href{https://doi.org/10.1103/PhysRevD.21.1445}{\emph{Phys. Rev. D}
  {\bfseries 21} (1980) 1445}.

\bibitem{Hartle:1981zt}
J.B.~Hartle and G.T.~Horowitz, \emph{{Ground State Expectation Value of the
  Metric in the 1/$N$ or Semiclassical Approximation to Quantum Gravity}},
  \href{https://doi.org/10.1103/PhysRevD.24.257}{\emph{Phys. Rev. D} {\bfseries
  24} (1981) 257}.

\bibitem{RandjbarDaemi:1981wd}
S.~Randjbar-Daemi, \emph{{Stability of the Minkowski Vacuum in the Renormalized
  Semiclassical Theory of Gravity}},
  \href{https://doi.org/10.1088/0305-4470/14/7/001}{\emph{J. Phys. A}
  {\bfseries 14} (1981) L229}.

\bibitem{Jordan:1987wd}
R.D.~Jordan, \emph{{Stability of Flat Space-time in Quantum Gravity}},
  \href{https://doi.org/10.1103/PhysRevD.36.3593}{\emph{Phys. Rev. D}
  {\bfseries 36} (1987) 3593}.

\bibitem{Suen:1988uf}
W.-M.~Suen, \emph{{The Stability of the Semiclassical Einstein Equation}},
  \href{https://doi.org/10.1103/PhysRevD.40.315}{\emph{Phys. Rev. D} {\bfseries
  40} (1989) 315}.

\bibitem{Suen:1989bg}
W.M.~Suen, \emph{{Minkowski Space-time Is Unstable in Semiclassical Gravity}},
  \href{https://doi.org/10.1103/PhysRevLett.62.2217}{\emph{Phys. Rev. Lett.}
  {\bfseries 62} (1989) 2217}.

\bibitem{Anderson:2002fk}
P.R.~Anderson, C.~Molina-Paris and E.~Mottola, \emph{{Linear response, validity
  of semiclassical gravity, and the stability of flat space}},
  \href{https://doi.org/10.1103/PhysRevD.67.024026}{\emph{Phys. Rev. D}
  {\bfseries 67} (2003) 024026}
  [\href{https://arxiv.org/abs/gr-qc/0209075}{{\ttfamily gr-qc/0209075}}].

\bibitem{Matsui:2019tlf}
H.~Matsui and N.~Watamura, \emph{{Quantum Spacetime Instability and Breakdown
  of Semiclassical Gravity}},
  \href{https://doi.org/10.1103/PhysRevD.101.025014}{\emph{Phys. Rev. D}
  {\bfseries 101} (2020) 025014}
  [\href{https://arxiv.org/abs/1910.02186}{{\ttfamily 1910.02186}}].

\bibitem{Simon:1991bm}
J.Z.~Simon, \emph{{No Starobinsky inflation from selfconsistent semiclassical
  gravity}}, \href{https://doi.org/10.1103/PhysRevD.45.1953}{\emph{Phys. Rev.
  D} {\bfseries 45} (1992) 1953}.

\bibitem{Parker:1993dk}
L.~Parker and J.Z.~Simon, \emph{{Einstein equation with quantum corrections
  reduced to second order}},
  \href{https://doi.org/10.1103/PhysRevD.47.1339}{\emph{Phys. Rev. D}
  {\bfseries 47} (1993) 1339}
  [\href{https://arxiv.org/abs/gr-qc/9211002}{{\ttfamily gr-qc/9211002}}].

\bibitem{Will:2005va}
C.M.~Will, \emph{{The Confrontation between general relativity and
  experiment}}, \href{https://doi.org/10.12942/lrr-2006-3}{\emph{Living Rev.
  Rel.} {\bfseries 9} (2006) 3}
  [\href{https://arxiv.org/abs/gr-qc/0510072}{{\ttfamily gr-qc/0510072}}].

\bibitem{Will:2014kxa}
C.M.~Will, \emph{{The Confrontation between General Relativity and
  Experiment}}, \href{https://doi.org/10.12942/lrr-2014-4}{\emph{Living Rev.
  Rel.} {\bfseries 17} (2014) 4}
  [\href{https://arxiv.org/abs/1403.7377}{{\ttfamily 1403.7377}}].

\bibitem{Leseduarte:1996ah}
S.~Leseduarte and A.~Romeo, \emph{{Complete zeta function approach to the
  electromagnetic Casimir effect for spheres and circles}},
  \href{https://doi.org/10.1006/aphy.1996.0101}{\emph{Annals Phys.} {\bfseries
  250} (1996) 448} [\href{https://arxiv.org/abs/hep-th/9605022}{{\ttfamily
  hep-th/9605022}}].

\bibitem{Leseduarte:1996xr}
S.~Leseduarte and A.~Romeo, \emph{{Influence of a magnetic fluxon on the vacuum
  energy of quantum fields confined by a bag}},
  \href{https://doi.org/10.1007/s002200050331}{\emph{Commun. Math. Phys.}
  {\bfseries 193} (1998) 317}
  [\href{https://arxiv.org/abs/hep-th/9612116}{{\ttfamily hep-th/9612116}}].

\bibitem{Kay:1978zr}
B.S.~Kay, \emph{{The Casimir Effect Without {MAGIC}}},
  \href{https://doi.org/10.1103/PhysRevD.20.3052}{\emph{Phys. Rev. D}
  {\bfseries 20} (1979) 3052}.

\bibitem{Ferrara:1979wa}
S.~Ferrara, L.~Girardello and F.~Palumbo, \emph{{A General Mass Formula in
  Broken Supersymmetry}},
  \href{https://doi.org/10.1103/PhysRevD.20.403}{\emph{Phys. Rev. D} {\bfseries
  20} (1979) 403}.

\bibitem{Boyer:1968uf}
T.H.~Boyer, \emph{{Quantum electromagnetic zero point energy of a conducting
  spherical shell and the Casimir model for a charged particle}},
  \href{https://doi.org/10.1103/PhysRev.174.1764}{\emph{Phys. Rev.} {\bfseries
  174} (1968) 1764}.

\bibitem{Mahajan:2006mw}
G.~Mahajan, S.~Sarkar and T.~Padmanabhan, \emph{{Casimir Effect confronts
  Cosmological Constant}},
  \href{https://doi.org/10.1016/j.physletb.2006.08.026}{\emph{Phys. Lett. B}
  {\bfseries 641} (2006) 6}
  [\href{https://arxiv.org/abs/astro-ph/0604265}{{\ttfamily
  astro-ph/0604265}}].

\bibitem{Saghian:2012zy}
R.~Saghian, M.A.~Valuyan, A.~Seyedzahedi and S.S.~Gousheh, \emph{{Casimir
  Energy For a Massive Dirac Field in One Spatial Dimension: A Direct
  Approach}}, \href{https://doi.org/10.1142/S0217751X12500388}{\emph{Int. J.
  Mod. Phys. A} {\bfseries 27} (2012) 1250038}
  [\href{https://arxiv.org/abs/1204.3181}{{\ttfamily 1204.3181}}].

\end{thebibliography}\endgroup
\bibliographystyle{JHEP}

\end{document}